# A Particle Model That Produces Feynman Diagrams: Re-examination of Fundamental Entities, Free Particles, and Background Reference Frames

Marcia J. King
*228 Buckingham Avenue, Syracuse, NY 13210 and P. O. Box 1483, Borrego Springs, Ca 92004;*
*e-mail: mking52701@aol.com*

A relativistic quantized particle model avoids difficulties through (1) a Hamiltonian undecomposable into $H=H_0+H_I$, (2) a separation of the evolution parameter s from dynamics, (3) "leptons" and "hadrons" composed of "quarks," and (4) the absence of background reference frames. The stringlike Lagrangian is $L=-[-\mathbf{F}^2(\mathbf{Q})(d\mathbf{Q}/ds)^2+(\mathbf{F}d\mathbf{Q}/ds)^2]^{1/2}$. $\mathbf{Q}(s)$ defines quark positions; the form of $\mathbf{F}(\mathbf{Q})$ determines the interaction. The "strong" Lagrangian is symmetric under quark exchange. Transformation to new quark coordinates "hides" the symmetry. A variational principle for the parametrically invariant action in terms of these coordinates supplies natural boundary conditions (n.b.c.). The resulting symmetry breaking yields "lepton" and "hadron" quarks that behave differently. However, both become "strings" asymptotically. The n.b.c. produce composite mass-shell constraints and suppress time-oscillations. "Strong" scattering between composites is calculated. The "leptons" behave as free particles. A second choice of $\mathbf{F}(\mathbf{Q})$ produces unified "electroweak" interactions. First-order perturbation theory is applied to "lepton-lepton" scattering. Unperturbed states are asymptotic solutions from separate "strong interaction" clusters. Transforms between position and momentum representations, determined by the n.b.c., eliminate advanced potentials. Scattering amplitudes obey Feynman rules.

## I. INTRODUCTION

In this paper, a model, not a theory, is presented. It has a particle ontology yet shares characteristics with both quantum field theory and string theory. Although it derives from the research of recent decades, it necessarily demands a simultaneous re-examination of several basic ideas underlying current particle physics: (1) The use of physically extended objects as fundamental entities (namely, fields and strings), (2) the role of free particles in formulating fundamental theories, (3) the nature of the fundamental constituents of leptons and hadrons, (4) the choice of evolution parameter, and (5) the use of background frames of reference. In regard to (1), quantum field theory is currently being questioned as a fundamental theory,[1,2,3,4,5] and some physicists are skeptical whether strings can be considered as fundamental.[3,6] Concerning (2), it has been argued by S. Weinberg that, with some caveats, "quantum mechanics plus Lorentz invariance plus cluster decomposition implies quantum field theory."[7] One of those caveats is that the Hamiltonian has a meaningful decomposition $H=H_0+H_I$. For the basic particle "quark" model of this paper, there is no such decomposition. The departure from (3) is connected to the concept of supersymmetry, for we assume that leptons as well as hadrons are composed of quarks, and we begin with a Lagrangian symmetric under the interchange of "lepton" and "hadron" quarks (when spin is introduced into this simple model, "lepton" and "hadron" composites take on half-integer and integer spin, respectively). Let us immediately reassure the reader that the model leads to composite "leptons" that behave like free point particles under the "strong" interactions.

The departure from idea (4) also involves a deep conceptual shift. In relativity, time is on an equal footing with space, but in nonrelativistic quantum mechanics, time t is treated as a c-number and separated from the dynamical variables. On the other hand, canonical methods of quantization are based on nonrelativistic quantum mechanics. This discrepancy, known from the inception of quantum mechanics, was reviewed by K. Kuchar[8] in regard to attempts to apply canonical methods, including Dirac's constraint method, in quantizing relativistic theories. One of Kuchar's suggestions: Put the dynamical variables and time on an equal footing. In the present model, this is accomplished by expressing all four-vector variables in terms of an (unmeasurable) evolution parameter s. Thus, the kinematics (evolution) is separated from the dynamics. In doing this, we must recognize that Poincaré invariance now implies conservation laws in the variable s, and examine the conditions under which the usual space-time conservation laws hold in an observer's time. It is found that these conditions are closely linked to the particle-antiparticle concept.

In regard to (5), the basic model is formulated in the absence of a background frame of reference and thus may have implications for formulations of cosmological quantum theory. A. Ashtekar and J. Lewandowski[9] and C. Rovelli[10] are among those who emphasize the necessity of doing quantum physics in the absence of a space-time geometry. The elimination of background reference frames was also considered in the past by J. B. Barbour and L. Smolin.[11]

In spite of the differences with mainstream ideas discussed above, the basic assumptions underlying the model of this paper bear a resemblance to much that is current in particle physics. To begin with, the construction of the Lagrangian parallels gauge theory in the following way. It is assumed that the fundamental Lagrangian must exhibit a "spontaneously broken" internal symmetry (in this case, in this simplified model, it is a symmetry between two systems of quarks). In other words, without the addition of any other conditions "by hand," the symmetry must be such that the solutions, as a class, do not exhibit that symmetry of the Lagrangian.[12] The analogy to gauge theory continues: Under a transformation of coordinates, the symmetry becomes "hidden" while, at the same time, composite-particle mass shell constraints $p^2=m^2$ are "revealed."

A second similarity to current physics comes from the "string-like" form of the Lagrangian. It is assumed that a general Lagrangian holds for *all* interactions and has the form

$$L = -\left\{ -\left[\mathbf{F}\langle\mathbf{Q}(s)\rangle\right]^2 \dot{\mathbf{Q}}^2(s) + \left[\dot{\mathbf{Q}}(s)\cdot\mathbf{F}\langle\mathbf{Q}(s)\rangle\right]^2 \right\}^{1/2}. \tag{1.1}$$

$\mathbf{Q}(s)$ is a vector whose components are the position vectors of the "spinless" quarks of the system; $\mathbf{F}(\mathbf{Q})$ is a vector function; and s is the evolution parameter. The Lagrangian (1.1) is Lorentz covariant and leads to a parametrically invariant action. Presumably, a good choice of $\mathbf{F}(\mathbf{Q})$ would lead to a "true" unification model. Lacking this, we assume different functions for $\mathbf{F}(\mathbf{Q})$ for two kinds of interactions, the "strong" and the "electroweak." However, within the model the "electrodynamic" and "weak" interactions are truly unified. Moreover, we require that the solutions match at the boundaries of the "strong" and "electroweak" domains.

A third link with current physics is the assumption of a form of "supersymmetry." Although the quarks are assumed to be spinless, the assumed fundamental Lagrangian is symmetric under the interchange of two varieties of quarks. This symmetry is spontaneously broken, however, so that composite particles, which are observable (i.e., obey mass-shell constraints), are composed of pairs of either "lepton" or "hadron" quarks, rather than mixtures thereof. The behaviors of the resulting "leptons" and "hadrons"are very different under these "strong" interactions, with the "leptons" behaving, in essence, as free particles. In spite of this difference in behavior, the "lepton" and "hadrons" share the same mass spectrum. In Sec. VII, a brief consideration of spin variables suggests that the "leptons" have half-integer spin while the "hadrons" have integer spin.

The model itself is simple and the input very limited, but as the remarks above suggest, its basis is new and the resulting implications are broad in nature. Therefore, before presenting the details of the model, let us have a little closer look at the concepts involved.

### A. Why have fields and/or strings been considered necessary?

In recent years, concern has been expressed about the usefulness of quantum field theory in advancing our knowledge of fundamental processes.[1,2,3,4,5] There are many unanswered questions about problems such as quantum gravity, the true unification of forces, the calculation of testable predictions, and the interpretation of quantum mechanics. Although, for many, superstring theory holds much promise of providing a "new physics,"[4,5,13] there are those who question whether it will provide answers to such questions.[3,6] One purpose of this paper is to suggest that, in the light of new insights in particle physics, it may be worthwhile to re-examine old concepts. There were, for example, in the growingly distant past, attempts to construct relativistic quantum particle theories,[14] but this area of study largely disappeared before the discoveries of the recent decades.

The use of fields in formulating the description of dynamical systems has gone far beyond the classical field theories of Maxwell and Einstein.[15] In particular, quantum fields have become the basic ingredients in the formulation of relativistic particle physics. The concept of a field arose in physics partly to offer an intuitive explanation of how forces act over distance. This allowed the elimination of the conceptually difficult idea of forces transmitted through a vacuum. Although the idea of a void was considered by the ancient Greeks, it continues to puzzle us: Does nature exhibit bodies separated by "voids" or bodies immersed in some kind of "ether?" Since this question has not been settled, one has to ask why field theory has become so ubiquitous. Why not a relativistic particle mechanics? Although a good deal of attention has recently been given to the conceptual foundations of field theory, very little consideration has been given to former relativistic particle ontologies. Apparently, relativistic particle ontologies are not thought to be viable. S. Weinberg, for example, demonstrates that, with certain caveats, the only way to reconcile quantum mechanics with special relativity and cluster decomposition is through quantum field theory.[5,7]

Field theory did not immediately permeate all of particle physics. Difficulties, such as infinite self-energies, kept alive attempts to formulate relativistic particle dynamics throughout the middle decades of the 20th century.[13] The best known attempt is Wheeler and Feynman's action-at-a-distance model of electrodynamics.[16,17] However, all efforts failed in one way or another. These failures (for example, particles exceeding the velocity of light, lack of quantization, lack of cluster decomposition, and the inability to create and annihilate particles) contributed to the further expansion of field theory. The difficulties of infinities in field theory can be overcome, for example, by renormalization, even if the means are complicated.

The field theory approach in particle physics culminated with the formulation of the standard model. As extremely successful as it has been, it is not believed to be the final answer for several reasons.[18] One compelling reason is that the standard model does not represent a "true" unification of physical laws. A "true" unification implies that all the forces of nature are but different aspects of a single law.

It is the possibility of a true unification, one that includes gravity, in a unique theory that underlies the attraction of superstring theory.[13,19,20] Superstring theory is, at present, the only candidate for such a unification. It is also attractive because it is a theory with no free parameters. However, superstring theory suffers from the lack of physical predictions that can be tested experimentally. Moreover, the necessity of introducing extra unseen spatial dimensions on the basis of mathematical consistency alone has prompted criticism that the theory is unrealistic.[3,6]

## B. A source of complexity for field and string theories: Physically extended objects

The continuous field or string can be thought of as an infinite number of beads along an imaginary string. This produces difficulties because separate modes of oscillation must be considered for each "bead." It is the physical extensions of fields and strings that forbid their descriptions the simple elegance of nonrelativistic classical and quantum mechanics. Field theories are marred by the difficulties of mathematical divergences. Superstring theory is further complicated because the strings themselves are also physically extended objects. The algebra of the constraint functions cannot be closed without the introduction of extra spatial dimensions.

Dirac was concerned with preserving the successes of (nonrelativistic) classical mechanics during the development of quantum mechanics. He wrote,[21] "The underlying ideas and laws [of classical mechanics] . . . form a simple and elegant scheme, which one would be inclined to think could not be seriously modified without having all its attractive features spoilt." Due in part to Dirac's own work, the exposition of nonrelativistic quantum mechanics retains a good deal of the beauty of nonrelativistic classical physics.

The formulations of relativistic particle field theories and superstring theories represent two very successful and interesting developments in physics during the twentieth century. Through them, elegant new mathematical tools have been developed that may be useful in the future, and many insights have been gained about diverse problems such as spontaneous symmetry breaking, the inclusion of gravity within a quantum theory, the origins of the Universe, and the behavior of black holes. Nevertheless, one cannot say these formulations have the simplicity and elegance of either classical physics or nonrelativistic quantum mechanics.

## C. The roles of free particles and free fields

There is another entrenched idea (besides fields), basic to most of physics, which probably has its roots in the ancient Greek pre-occupation with individuality and freedom. In physics, from classical particle dynamics to quantum field theories, the equations for a free particle or field generally form the starting point, after which interactions are introduced. This is often characterized by writing the Lagrangian as a sum of the free and interaction parts, or L = T+V. The assumption in a different form, $H = H_O + H_I$, is the basis for an argument by S. Weinberg that quantum mechanics cannot be reconciled with special relativity.[5,7]

### D. A constituent model for which there is no decomposition $H=H_0+H_I$

In light of the difficulties of field and string theories, it is worthwhile to reconsider whether it is necessary to introduce physically extended objects as the basic components of any theory. The research of the last few decades gives us a fresh vantage point in the consideration of relativistic particle dynamics and the problems that were previously encountered. It is further intriguing, as we shall see, to consider the possibility of solving the earlier problems of particle dynamics by eliminating free particles as the basis of the theory.

Nevertheless, like Dirac, we would hate to lose the attractive features of current theories in choosing a new approach. Among the most intriguing aspects of field theory are the harmonic oscillator operators a(k) and $a^\dagger(k)$ which arise in the Fourier transform of the free field associated with a particular particle. Weinberg[22] has pointed out that quite apart from being annihilation and creation operators, these operators make possible cluster decomposition, i.e., the uncorrelated results of distant experiments.

The harmonic oscillator has played a role in various nonrelativistic quark models of hadrons, and was extended to include relativistic calculations by Feynman, Kislinger, and Ravndal.[23] However, the problem of time-oscillations prevented the nonrelativistic quark model from being rigorously extended to the relativistic domain. Thus, in considering a relativistic *particle* dynamics, we must first construct a model that can be considered realistic.

We shall see that the relativistic harmonic-oscillator Hamiltonian for the "strong" interaction version of this model cannot be broken up into $H = H_0 + H_I$ in any meaningful way. The basic assumptions underlying the model are (1) *all* observable particles consist of constituents we shall call "quarks," and (2) spontaneous symmetry breaking results in two varieties of quarks which will be termed "lepton" and "hadron" quarks. Concomitantly, the symmetry breaking also results in two kinds of observable composite particles, "lepton" and "hadrons." The possibility of bypassing the problem of "unphysical" behavior is quite clear. "Unphysical" quarks can be allowed provided they cannot be observed directly and the observable particles they make up behave "physically." Note that this was anticipated by the quark-exchange "duality diagrams" of dual models of the '70s in which quarks can be interpreted as turning around in time.[24] The quarks can considered as unobservable if, by assuming quark confinement in the initial, quarks are also confined in the final state. The unobservable quarks brings into question whether the ontology of the model is to be based on particles or structural relationships.[1,2] However, here we shall not discuss this deep philosophical question. If the quarks are not "real" particles in the usual sense, they are nonetheless a mechanism for obtaining "real" results.[25]

The assumption that all observable composites consist of quarks is basic to the model in all energies ranges, not just the “strong-interaction” domain. Leptons, of course, have not heretofore been considered to be composed of quarks. It may be worthwhile to point out an incongruity of the standard model. The assumed fundamental particles, namely quarks and leptons, have very different properties. Why should nature group them together? The results of this paper suggest that this may have been the stumbling block to true unification.

## E. The outline of the paper

A naive heuristic model is presented which suggests the following possibilities: (1) Constituent quarks allow a self-consistent formulation of quantum relativistic particle dynamics in which subsidiary conditions are built in as part of the dynamical system (natural boundary conditions); (2) quarks are the fundamental constituents of both hadrons and leptons; (3) these quarks yield a true unification of the electrodynamic and weak interactions; and (4) they lead to the derivation of Feynman rules for composite scattering in electroweak perturbation theory.

The model is kept simple in order to readily exhibit the essential features of the new concepts. Discussion is limited to composite particles whose constituents are pairs of spinless “quarks.”[26] Thus, the composite particles of the model cannot directly represent experimentally observed leptons and hadrons. Nevertheless, we shall take the liberty of choosing the same nomenclature, that is, just as we have chosen to call the constituent particles “quarks,” we shall also use the labels “leptons” and “hadrons” to describe the composite particles of the model. From now on, however, we shall omit the quotation marks. We shall also use the terms “strong” and “electroweak” interactions in the same manner. These loosely defined terms allow the model to be more intelligible and the language less cumbersome, but they should not be taken literally.

In Sec. II, we discuss the use of the principle of Least Action[27] associated with the integral

$$A = \int_{s_1}^{s_2} \sum_{i=1}^{n} p_i(s)\, \dot{q}_i(s)\, ds. \tag{1.2}$$

This is because, for a parametrically-invariant action, the action functional takes on the same form, or

$$I \equiv \int_{s_1}^{s_2} L(s)\, ds = \int_{s_1}^{s_2} \sum_{i=1}^{n} p_i(s)\, \dot{q}_i(s)\, ds. \tag{1.3}$$

The variation is taken to vanish identically, i.e., $\Delta I = 0$. Since the variations of the parameter s at the end points do not vanish,[27] this leads not only to Lagrange's equations but also to a set of natural boundary conditions[28] of the form

$$p_i^2(s_1) = 0, \qquad p_i^2(s_2) = 0, \quad i = 1, 2, 3, \ldots, n. \tag{1.4}$$

These natural boundary conditions (n.b.c.) result in a form of "spontaneous symmetry breaking," for they are a part of the dynamical system.[12] The Lagrange equations of motion are invariant under a larger group of transformations than the n.b.c., for example, contact transformations. Thus, the variational principle selects out special coordinate frames. We shall assume that these frames are correlated with the fact that we, the observers, are made of the same particles that we observe, and the variational principle is applied to the sets of coordinates which represent observable (composite) particles.

For the model in this paper, the n.b.c. yield the necessary subsidiary conditions to suppress unwanted time oscillations and provide mass-shell constraints for composite particles, that is, the Klein-Gordon equations $p^2 = m^2$. As we shall see in Sec. IV, the n.b.c. also prohibit composites consisting of mixtures of lepton and hadron quarks.

In Sec. III, a harmonic-oscillator potential is adopted. The Lagrangian in an arbitrary Lorentz frame is

$$L(s) = -M\left\{-[\mathbf{G}\mathbf{Q}(s)]^2\dot{\mathbf{Q}}^2(s) + [\dot{\mathbf{Q}}(s)\mathbf{G}\mathbf{Q}(s)]^2\right\}^{1/2}, \tag{1.5}$$

where M is a constant parameter and **G** is a numerical dimensionless matrix yielding a harmonic-oscillator interaction. The variable s is an evolution parameter. In the past, such parameters have sometimes been associated with proper time. We do not do so, and later find that s is unmeasurable. The variable **Q**(s) is a vector whose components are the individual quark position four-vectors. For simplicity, the quarks are taken to be spinless. The harmonic-oscillator coupling matrix **G** is chosen to describe two identical but noninteracting systems that we shall denote as the lepton and hadron quarks, respectively. Composite particles, i.e., leptons and hadrons, can be formed by appropriate initial conditions on quark solutions. The solutions imply a lepton-hadron symmetry we do not see in nature. Also, the composites in this version of the model do not interact among themselves. Therefore, we have two reasons to consider the symmetry as only approximate. Just as for quantum field theory, we modify the Lagrangian so that spontaneous symmetry breaking provides the interactions between particles as well as breaking unwanted symmetries.

In Sec. IV, the harmonic-oscillator potential of Sec. III is modified by a familiar prescription. An infinitesimal positive imaginary part, proportional to $i\varepsilon$, is added to the parameter M and the coupling matrix **G**. This leads to a new harmonic-oscillator quark model which is still characterized by a symmetry under the interchange of coordinates. However, these are transformed to a new set of coordinates designated as the quark coordinates and yields the Lagrangian

$$L = -M\left\{ -\left[\mathbf{HQ}(s)\right]^2 \dot{\mathbf{Q}}^2(s) + \left[\dot{\mathbf{Q}}(s)\,\mathbf{HQ}(s)\right]^2 \right\}^{1/2} . \tag{1.6}$$

The original symmetry becomes hidden: The coupling matrix **H** is not symmetric under the interchange of "lepton" and "hadron" quarks**.** The principle of least action applied to (1.6) yields a set of natural boundary conditions (n.b.c.) that suppress unwanted time oscillations and ensure the mass-shell relations $p^2=m^2$ for composite particles. It also restricts composites to leptons and hadrons (no "mixed" states). Quark masses have no meaning in this model, except in the asymptotic regions of s, where they could be considered as being zero (see (1.4)). Curiously, the quarks also become "strings" in those asymptotic regions (their frequencies of oscillation tend to infinity as $\varepsilon \to 0$).

A set of Dirac constraints results from this string-like Lagrangian that are used to construct a Dirac Hamiltonian[29] and thus a quantized theory. The Hamiltonian is characterized by an arbitrary gauge parameter which we take to be inversely proportional to $\varepsilon$. For the strong interaction Lagrangian, the constraints are automatically satisfied by the n.b.c. (the converse is not true). The general formalism is discussed in Sec. IV, and the equations of motion are solved. The n.b.c., which are part of the dynamical system, imply a reformulation of the usual completeness relations. The nature of the solutions of (1.6) lead to new transformations between dynamically allowed states in the position and momentum representations.

The harmonic-oscillator potential allows cluster decomposition, i.e., distant kinematically uncorrelated systems of unrestricted numbers of quarks, and thus any number of virtual particles. In the interest of simplicity, the clusters considered in this paper are limited to systems of four quarks. These, in turn, lead to the elastic scattering of like composite particles by quark exchange.

Feynman[30] demonstrated that quantum mechanics and special relativity (in the sense of restricting the velocity of observable particles to less than c) requires the existence of antiparticles. Here, we have the situation that the quark velocities are not restricted, although the velocities of the observable composites are. Nevertheless, in Sec. IV we define antiquarks in analogy to antiparticles. Quark-antiquark confinement appears as part of the physical initial conditions in the harmonic-oscillator model, but once imposed in the observer's initial state, follows also in the final state.

The strong interaction scattering of like composites is considered in Sec. V. A new interpretation of the state vector $|\Psi\rangle$ arises and must be considered when calculating scattering amplitudes. It also impacts on the interpretation of quantum mechanics. The relativistic state vector $|\Psi\rangle$ can be interpreted as an extension of the nonrelativistic time-independent state vector $|\Psi(\mathbf{x})\rangle$ which contains information for all values of **x**. In the present case, the parametric invariance of the Lagrangian implies there is no evolution of the system in the parameter s (because of the vanishing of the Hamiltonian). Nevertheless, the state vectors for the system contain complete information about the particles at the observer's times $t = \pm\infty$. Thus, the states cannot be completely specified by the physical initial conditions at $t = -\infty$. In fact, even both initial and final physical measurements together do not specify the state vector completely. The remaining parameters must be summed over in calculating the overlap of possible initial and final states.

The lepton composites can scatter only in the forward or backward directions, which for these identical particles is tantamount to no scattering at all. The hadrons scatter at other angles as well. Hadron-hadron scattering amplitudes are calculated and take a "Veneziano" type form characteristic of dual models:[24]

$$A(s,t,u) = \prod_{I=1}^{4} \delta(P_I^{\,2} - M^2)\ \delta(P_1 + P_2 + P_3 + P_4)\ [D(s) + D(t) + D(u)], \tag{1.7}$$

where

$$D(z) = \sum_{n=-\infty}^{\infty} \delta(z-n) = (2\pi i)^{-1} \sum_{n=-\infty}^{\infty} \left[ \frac{1}{z-n-i\varepsilon} - \frac{1}{z-n+i\varepsilon} \right]. \tag{1.8}$$

The orbital angular momentum is always zero in this simple model (discussed shortly), resulting in constant residues for poles in s, t, and u.

It will be demonstrated in Sec. V that the scattered composite particles carry no memory of the interaction, save for the usual conservation laws of four-momentum and angular momentum. Thus composites from different (uncorrelated) clusters may be used to provide the basis states for the unperturbed Hamiltonian in the electroweak domain.

Now let us add some comments about the origin of (1) the absence of background reference frame, (2) no decomposition $H=H_0+H_I$, and (3) the zero orbital angular momentum of the composites. These all stem from the addition of the infinitesimal parameter $\varepsilon$ to the harmonic-oscillator model, one result of which is that the quark position vectors are singular as $\varepsilon \to 0$. (Matrix elements are always calculated before going to the limit $\varepsilon = 0$.) In compensation, the volume in space-time within which the interaction takes place (defined by finite values of s) goes to zero as $\varepsilon \to 0$. This also implies a direct collision of the composites.

The singular nature of the quark coordinates plays an important role. The strong scattering of like composites takes place in a special coordinate frame (although Lorentz covariance is maintained), namely the frame in which the variational principle is applied. (It is in this frame, for this model, that the quark momenta and position vectors are aligned, i.e., the composites have zero orbital angular momentum.) On the other hand, the original fundamental Lagrangian, obtained by modifying the lepton-hadron symmetric Lagrangian of Sec. III, is expressed in coordinates for which the total four-momentum and angular momentum of the quark system vanishes. Furthermore, the total center-of-mass vector plays no role in the scattering of observable composite particles. As a result, the composite coordinates, both external and internal, are all relational. *There is no background frame of reference.* This has an important significance when final states representing free composite particles are used as the unperturbed basis states for subsequent interactions. We shall see in Sec. V that, in combination the role of the observer and the n.b.c., it implies a modification of the transformations between momentum and position representations. One result is the exclusion of "advanced" electroweak potentials which, in turn, allows the subsequent derivation of Feynman propagators and Feynman diagrams.

In Sec. VI, the electroweak interactions are considered. The Lagrangian becomes

$$L(s) = -\{-[\mathbf{A}(\mathbf{Q}(s))]^2 \dot{\mathbf{Q}}^2(s) + [\mathbf{A}(\mathbf{Q}(s)) \cdot \dot{\mathbf{Q}}(s)]^2\}^{1/2}. \tag{1.9}$$

In choosing an expression for $\mathbf{A}(\mathbf{Q})$, one's first impulse is to borrow from the classical Wheeler-Feynman model of action-at-a-distance electrodynamics.[16,17] However, it appears that a Wheeler-Feynman type potential can yield Feynman's rules for "electromagnetic" but not "weak" scattering amplitudes. Interestingly, as is shown, the Wheeler-Feynman potential can be regarded as a four-dimensional generalization of a simple one-dimensional problem, namely, the nonrelativistic collision of two point particles. This one-dimensional problem can be formulated in other ways. We generalize the simplest of these alternatives to four-dimensional space-time because it leads to a "true" unification of the "electrodynamic" and the "weak" interactions which yields Feynman diagrams.

First-order perturbation theory is applied to an example of spinless lepton-lepton elastic scattering (the analogous hadron-hadron scattering can be similarly formulated). The unperturbed states are assumed to be asymptotic harmonic-oscillator lepton states generated from the harmonic-oscillator domains. With the aid of the new completeness relations, the Dirac propagator $\Delta_F(x)$ for the Klein-Gordon equation is derived, and Feynman rules follow. There are no advanced potentials.

Finally, in order to make a connection to the usual understanding of supersymmetry, internal spin variables are briefly discussed in Sec. VII. A second spin phase space is added to the phase space of quark particles. The composite particles in this model are composed of pairs of quarks, leading to leptons and hadrons with half-integral and integral spin, respectively. The leptons and hadrons share the same mass spectrum.

## II. VARIATIONAL PRINCIPLE AND NATURAL BOUNDARY CONDITIONS

Consider a system of n particles described by coordinates $q_i(s)$ and velocities $\dot{q}_i(s) = dq_i(s)/ds$, with i = 1, 2, 3, . . . , n. The variable s is an “evolution” parameter. Assume the Lagrangian L is a known function of $q_i(s)$ and $\dot{q}_i(s)$ and is nonstandard. That is, the canonical Hamiltonian vanishes, or

$$L\left[q_i(s), \dot{q}_i(s)\right] - \sum_{i=1}^{n} p_i(s)\,\dot{q}_i(s) = 0, \tag{2.1}$$

where $p_i \equiv \partial L/\partial \dot{q}_i$. Then the action functional can be defined as

$$I[C] = \int_{s_1}^{s_2} ds\, L\left[q_i(s), \dot{q}_i(s)\right] = \int_{s_1}^{s_2} ds \sum_{i=1}^{n} p_i(s)\,\dot{q}_i(s)\,. \tag{2.2}$$

Now consider a path $C'$ differing infinitesimally from C and calculate the variation

$$\Delta I \equiv I[C'] - I[C]. \tag{2.3}$$

The $\Delta$ variation needs to be specified. In view of (2.1) and (2.2), we shall adopt the variational procedure used for a different integral associated with standard Lagrangians and defined as

$$A \equiv \int_{s_1}^{s_2} \sum_i p_i(s)\,\dot{q}_i(s)\,ds, \tag{2.4}$$

The variational principle is sometimes known as *the principle of Least Action.*[27] The displacements of $q_i(s)$ and $\dot{q}_i(s)$ are affected only by the “speeding up” or “slowing down” of each variable as a function of s, leaving the varied path consistent with the physical motion. Thus, the $\Delta$-process includes a variation of s even at the end points, but the variation of the $q_i$’s remain zero. We consider the variations

$$q_i'(s) \equiv q_i(s) + \delta_i\, q_i(s), \tag{2.5}$$

where

$$\delta_i q_i(s) \equiv \dot{q}_i(s)\,\delta_i s. \tag{2.6}$$

We have also

$$\dot{q}_i'(s) \equiv \dot{q}_i(s) + \frac{d}{ds}\delta_i q_i(s). \tag{2.7}$$

Now compute the first-order change in the action functional I in going from C to C′:

$$\Delta I \equiv I[C'] - I[C] = \int_{s_1}^{s_2} L\left[q_i'(s), \dot{q}_i'(s)\right] ds - \int_{s_1}^{s_2} L\left[q_i(s), \dot{q}_i(s)\right] ds$$

$$= \int_{s_1}^{s_2} ds \sum_i \left[\frac{\partial L}{\partial q_i}\delta q_i(s) + \frac{\partial L}{\partial \dot{q}_i}\frac{d}{ds}\delta q_i(s)\right]$$

$$= \int_{s_1}^{s_2} ds \left[\sum_i \frac{\partial L}{\partial q_i} - \frac{d}{ds}\left(\frac{\partial L}{\partial \dot{q}_i}\right)\right]\dot{q}_i(s)\,\delta_i s + \sum_i \frac{\partial L}{\partial \dot{q}_i}\dot{q}_i(s)\,\delta_i s\,\Big|_{s_1}^{s_2}. \tag{2.8}$$

The last step was obtained by integrating by parts.

We now assume the total variation vanishes,[31] i.e.,

$$\Delta I = 0. \tag{2.9}$$

The variations $\delta_i s$ are taken to be arbitrary and independent, and the end points $s_1$ and $s_2$ are arbitrary as well. Thus, for nontrivial solutions (where the $\dot{q}_i(s)$ are not identically zero), it follows that

$$\frac{\partial L}{\partial q_i} - \frac{d}{ds}\left(\frac{\partial L}{\partial \dot{q}_i}\right) = 0, \tag{2.10}$$

and

$$p_i(s)\,\dot{q}_i(s)\big|_{s_1} = 0, \qquad p_i(s)\,\dot{q}_i(s)\big|_{s_2} = 0. \tag{2.11}$$

The last two conditions are known as the *natural boundary conditions*.[28] We shall denote them as the n.b.c.

The form of the Lagrange equations (2.10) is invariant under contact transformations. That is, for canonical transformations to a new set of coordinates $Q_i$ and $P_i$ such that

$$\sum_i P_i dQ_i = \sum_i p_i dq_i. \tag{2.12}$$

However, the n.b.c. are not form invariant under such transformations. Thus, the variational principle selects out special coordinate systems. For quark solutions in the harmonic-oscillator model of Secs. III and IV, the n.b.c. take the form $p_i^2=0$. Then, at least, the n.b.c. are invariant under the Lorentz transformations. To see what kind of "special" frame might be involved here, let us consider the choices of coordinates employed in nonrelativistic classical mechanics for standard Lagrangians. For a system of n particles, we assign, *on the basis of what we observe (measure),* a set of "coordinates" $x_i(t)$ and their derivatives $\dot{x}(t) = dx/dt$. We might then choose, for reasons of convenience in solving equations, say, to make a transformation to a different set of coordinates, to which we, of course, don't assign a particle interpretation. Thus arises the concept of "generalized" coordinates. The variation principle assumed for standard Lagrangians includes the assumption of no variation at the end points, and therefore gives equivalent descriptions in the two cases: It does not select out which set of coordinates correspond to the "real" particle coordinates.

For the nonstandard Lagrangians, let us then assume that the variational principle described in this section is to be applied only to Lagrangians expressed in terms of real particle coordinates and their derivatives. In other words, *it is assumed that the variational principle is based on a particle ontology*.

We must now add a caveat, however. The above assumption is based on the fact that we, by the act of observation (measurement) necessarily have a built-in bias. By all appearances, we and our measuring devices consist of the same kinds of particles we measure. This may imply that we are incapable of direct measurement of other kinds of "particles," but it doesn't preclude their existence. As we shall see, the results of the model in the present paper re-enforce that viewpoint: We find that observable particles "in" imply observable particles "out," but the existence of other kinds of particles are part of the model. Thus, we shall assume that the Lagrangian in terms of "measurable particle" coordinates is not necessarily the "fundamental" Lagrangian.

This version of the variational principle has not been adopted before for nonstandard Lagrangians in particle physics, and, even without considering the lack of form invariance under contact transformations, it's not hard to see why. If it is applied for "ordinary" relativistic particles, it can lead to inconsistencies. For example, consider the relativistic Lagrangian for a free particle (setting the mass equal to unity):

$$L = \left[\dot{x}^2(s)\right]^{1/2}. \tag{2.13}$$

The conjugate momentum is

$$p = \frac{\dot{x}(s)}{\left[\dot{x}^2(s)\right]^{1/2}}, \tag{2.14}$$

which yields the constraint

$$p^2 = 1. \tag{2.15}$$

This is in contradiction with the n.b.c. (Recall the comments in the Introduction concerning the assumptions of free particles in theory construction.) We shall see in the following sections that, for quark coordinates, such inconsistencies do not arise. In fact, the n.b.c. for our relativistic model produce the orthogonality conditions necessary to ensure the suppression of unwanted time oscillations and composite particle mass-shell constraints $p^2=m^2$.

Not only is there a close connection between the choice of frame for application of the variational principle and observable particles, but there is a primary significance to be considered once the system is quantized. That is because the n.b.c. are a part of the dynamical system. The complete set of solutions for the equations of motion are not the complete set of solutions for the dynamical system. Thus, it is necessary to revise formulations involving complete sets of states ("completeness relations").

## III. THE FIRST STEP: A LEPTON-HADRON SYMMETRIC QUARK MODEL

### A. Nonstandard "stringlike" Lagrangian and corresponding Dirac Hamiltonian

Consider a system of 4N quarks described by coordinates $Q_{IA\mu}(s)$ and $\dot{Q}_{IA\mu}(s) \equiv dQ_{IA\mu}(s)/ds$, where I=1, 2, 3, ... , N, and A=1, 2, 3, 4. For simplicity, the quarks are assumed to be spinless. The parameter s is an evolution parameter. The metric is $g_{00}=1 = -g_{ii}$. We assume a nonstandard Lagrangian in an arbitrary Lorentz frame which leaves the action parametrically invariant:

$$L(s) = -\left\{M^2\left[-\left[\mathbf{GQ}(s)\right]^2 \dot{\mathbf{Q}}^2(\mathbf{s}) + \left[\dot{\mathbf{Q}}(s)\cdot\mathbf{GQ}(s)\right]^2\right]\right\}^{1/2}. \tag{3.1}$$

We have set $\hbar = c = 1$, M is a constant parameter, and

$$\mathbf{Q} = \begin{bmatrix} Q_{11} \\ Q_{12} \\ \cdot \\ \cdot \\ \cdot \\ Q_{N4} \end{bmatrix}. \tag{3.2}$$

The coupling matrix **G** is given by

$$\mathbf{G} \equiv \mathbf{g} \otimes \mathcal{N}, \tag{3.3}$$

with the definitions

$$\mathbf{g} \equiv \frac{1}{2}\begin{bmatrix} 1 & 0 & -1 & 0 \\ 0 & 1 & 0 & -1 \\ -1 & 0 & 1 & 0 \\ 0 & -1 & 0 & 1 \end{bmatrix} = \mathbf{g}^2, \quad \text{and} \quad \mathcal{N} \equiv \frac{1}{N}\begin{bmatrix} 1 & 1 & 1 & . & . & . & 1 \\ 1 & 1 & 1 & . & . & . & 1 \\ 1 & 1 & 1 & . & . & . & 1 \\ . & . & . & . & . & . & . \\ . & . & . & . & . & . & . \\ . & . & . & . & . & . & . \\ 1 & 1 & 1 & . & . & . & 1 \end{bmatrix} = \mathcal{N}^2. \tag{3.4}$$

Note that **G** = **G**$^2$.

The momentum conjugate to **Q** is

$$\mathbf{P} = \frac{\partial L}{\partial \dot{\mathbf{Q}}} = \frac{M^2}{L}\left[-(\mathbf{GQ})^2\dot{\mathbf{Q}} + (\dot{\mathbf{Q}}\cdot\mathbf{GQ})\mathbf{GQ}\right]. \tag{3.5}$$

The first term under the square root in the Lagrangian (3.1) is usually identified as VT, or the potential times the kinetic energy. Models closely similar to (3.1) and the broken symmetry Lagrangian of the next section are considered in Reference 26. In these earlier models, the Lagrangians took the form $L = -\sqrt{VT}$, and in that case the only Dirac constraint is $H \approx 0$.

Expanding the potential term in (3.1), we obtain the familiar harmonic oscillator form

$$V \equiv (M\mathbf{GQ})^2 = -\frac{M^2}{N}\sum_{I=1}^{N}\sum_{J=1}^{N}\sum_{A=1}^{4}\sum_{B=1}^{4} g^2{}_{AB}\left[Q_{IA}(s) - Q_{JB}(s)\right]^2. \tag{3.6}$$

We shall denote the A=1 and A=3 quarks as lepton quarks, and the A=2 and A=4 quarks as hadron quarks. Examination of the matrix **g** then shows that L describes two identical but independent systems. In other words, L remains unchanged under the simultaneous interchanges $Q_{I1} \leftrightarrow Q_{I3}$, $Q_{I2} \leftrightarrow Q_{I4}$. Thus this symmetry will be referred to as lepton-hadron quark symmetry.

The Lagrangian (3.1) bears a striking resemblance to the Lagrangian for a free string:[24]

$$L(\tau,\sigma) = -\left\{-x'^2(\tau,\sigma)\,\dot{x}^2(\tau,\sigma) + \left[\dot{x}(\tau,\sigma)\cdot x'(\tau,\sigma)\right]^2\right\}^{1/2}, \tag{3.7}$$

where x is a function of proper time $\tau$ and string parameter $\sigma$, $0 \le \sigma \le \sigma_0$. The derivatives are defined as

$$\dot{x}(\tau,\sigma) \equiv dx(\tau,\sigma)/d\tau; \quad x'(\tau,\sigma) = dx(\tau,\sigma)/d\sigma. \tag{3.8}$$

The analogy to the free string is discussed at length in the first paper of Ref. 26.

Returning to the quark model, we observe that the nonstandard L in (3.1) implies the vanishing of the canonical Hamiltonian. We turn to Dirac's theory of constraints[29] to construct a Hamiltonian. The Lagrangian (3.1) gives rise to two primary constraints:

$$\Phi_1 \equiv \mathbf{P}^2 + M^2\mathbf{QG}^2\mathbf{Q} \approx 0, \qquad \Phi_2 \equiv \mathbf{PGQ} \approx 0. \tag{3.9}$$

These may be used to construct a tentative Hamiltonian, $H = v_1\Phi_1 + v_2\Phi_2$, but we must now consider questions of consistency.

The equation of motion for a function f(Q,P) is given by

$$\dot{f} \approx \{f, H\} \tag{3.10}$$

with the Poisson brackets defined as

$$\{f,H\} \equiv \frac{\partial f}{\partial Q}\frac{\partial H}{\partial P} - \frac{\partial f}{\partial P}\frac{\partial H}{\partial Q}. \tag{3.11}$$

Taking f to be $\Phi_2$, we know that $\dot{\Phi}_2$ must be zero for consistency. We set

$$\{\Phi_1, \Phi_2\} = 2\Phi_3 \approx 0, \tag{3.12}$$

where we have defined

$$\Phi_3 \equiv \mathbf{PGP} - M^2\mathbf{QGQ}, \tag{3.13}$$

and we have used $\mathbf{G}^3 = \mathbf{G}$. Thus, we obtain the secondary constraint $\Phi_3 \approx 0$.

The Hamiltonian is now modified to be

$$H = \sum_{i=1}^{3} v_i\Phi_i \tag{3.14}$$

Once more, consistency requires

$$\{\Phi_i, H\} \approx 0, \quad i = 1,2,3. \tag{3.15}$$

This yields the conditions

$$v_2\left[\mathbf{PGP} + M^2\mathbf{QGQ}\right] \approx 0.$$

$$v_3\left[\mathbf{PGP} + M^2\mathbf{QGQ}\right] \approx 0, \tag{3.16}$$

Thus, to avoid adding a further constraint, set $v_2 = v_3 = 0$. Choosing a gauge $v_1 = \omega/(2M)$ = constant, we have

$$H = \frac{\omega}{2M}[\,\mathbf{P}^2 + M^2\mathbf{Q}\mathbf{G}^2\mathbf{Q}\,]. \tag{3.17}$$

The equations of motion $\dot{\mathbf{Q}} = \partial H/\partial \mathbf{P}$ and $\dot{\mathbf{P}} = -\partial H/\partial \mathbf{Q}$ yield

$$\ddot{\mathbf{Q}} = -\omega^2\,\mathbf{G}^2\mathbf{Q} \quad \text{and} \quad \mathbf{P} = (M/\omega)\,\dot{\mathbf{Q}}\ . \tag{3.18}$$

In order to solve these equations, it is convenient to transform to a set of "normal" coordinates. To this end, we note that the matrix **g** can be diagonalized by either of the following matrices $\delta$ or $\rho$:

$$\delta \equiv \frac{1}{\sqrt{2}}\begin{pmatrix} 1 & 0 & 1 & 0 \\ 0 & 1 & 0 & 1 \\ 1 & 0 & -1 & 0 \\ 0 & 1 & 0 & -1 \end{pmatrix} = \delta^{-1}, \qquad \rho \equiv \frac{1}{2}\begin{pmatrix} 1 & 1 & 1 & 1 \\ 1 & -1 & 1 & -1 \\ 1 & 1 & -1 & -1 \\ 1 & -1 & -1 & 1 \end{pmatrix} = \rho^{-1}. \tag{3.19}$$

It follows that

$$\delta^{-1}\mathbf{g}\delta = \rho^{-1}\mathbf{g}\rho = \begin{pmatrix} 0 & 0 & 0 & 0 \\ 0 & 0 & 0 & 0 \\ 0 & 0 & 1 & 0 \\ 0 & 0 & 0 & 1 \end{pmatrix}. \tag{3.20}$$

We shall work with the matrix $\delta$ (it is easily shown that $\rho$ leads to equivalent results). Define a transformation to a set of 4N coordinates:

$$Q_{IA} \equiv y_{IA} + N^{-1/2}\{\delta\mathbf{W}\}_A, \qquad \text{with } \sum_{I=1}^{N} y_{IA} = 0. \tag{3.21}$$

The latter condition brings the number of coordinates back to 4N.

In terms of the transformed coordinates, the conjugate momentum in (3.18) becomes

$$P_{IA} = (M/\omega)\,\dot{Q}_{IA} = (M/\omega)\left[\dot{y}_{IA} + N^{-1/2}\{\delta\dot{\mathbf{W}}\}_A\right]. \tag{3.22}$$

This suggests defining momenta corresponding to $y_{IA}$ and $W_A$ by

$$p_{IA} \equiv (M/\omega)\, \dot{y}_{IA}, \quad \text{with} \quad \sum_{I=1}^{N} p_{IA} = 0; \tag{3.23}$$

and

$$p_A \equiv (M/\omega)\, \dot{W}_A. \tag{3.24}$$

Then, we can write

$$P_{IA} = p_{IA} + N^{-1/2}\{\delta \mathbf{p}\}_A. \tag{3.25}$$

The Hamiltonian (3.17) can now be expressed as

$$H = \frac{\omega}{2M}\left[\sum_{I=1}^{N}\sum_{A=1}^{4} p_{IA}{}^2 + \sum_{A=1,2} p_A{}^2 + \sum_{A=3,4}\left(p_A{}^2 + M^2 W_A{}^2\right)\right]. \tag{3.26}$$

The variables $p_{IA}$ and $p_A$ are, in fact, conjugate to $y_A$ and $W_A$, respectively. The Hamiltonian or the equations (3.18) can be used to calculate the equations of motion

$$\ddot{W}_{IA} = 0; \qquad \ddot{W}_A = 0, \quad A = 1,\ 2; \qquad \ddot{W}_A = -\,\omega^2 W_A, \quad A = 3,\ 4. \tag{3.27}$$

## B. Quantized Solutions

We shall quantize by imposing the commutation relations

$$\left[y_{IA\mu}, p_{IA\nu}\right] = -\,i\, g_{\mu\nu}, \quad I = 1,\ 2,\ 3,\ \ldots,\ N-1;$$

$$\left[W_{A\mu}, p_{A\nu}\right] = -\,i\, g_{\mu\nu}, \quad A = 1,\ 2,\ 3,\ 4. \tag{3.28}$$

(Note that if we let index I run from 1 to N, there would be a contradiction. However, we can exclude the coordinates $y_{NA}$ and $p_{NA}$ by using (3.21) and (3.23).)

In the Heisenberg Picture, the equations of motion $\dot{O} = -\,i[O,H]$ yield equations (3.27). Express the solutions as follows:

$$y_{IA} = A_{IA}s + B_{IA}, \qquad p_{IA} = (M/\omega)\, A_{IA};$$

$$W_A = A_A s + B_A, \qquad p_A = (M/\omega) A_A, \quad A = 1,\ 2:$$

$$W_A = (2M)^{-1/2}\left[a_A{}^{\dagger} \exp(i\omega s) + a_A \exp(-\,i\omega s)\right],$$

$$p_A = i\,(M/2)^{1/2}\left[a_A{}^{\dagger} \exp(i\omega s) - a_A \exp(-\,i\omega s)\right], \quad A = 3,\ 4. \tag{3.29}$$

The commutation relations (3.28) imply

$$\left[a_{A\mu}, a^{\dagger}{}_{A\nu}\right] = -\,g_{\mu\nu}. \tag{3.30}$$

The Hamiltonian (3.26) in terms of the solutions (3.29) becomes

$$H = \frac{\omega}{2}\left[\frac{M}{\omega^2}\left\{\sum_{I=1}^{N}\sum_{A=1}^{4} A_{IA}{}^2 + \sum_{A=1}^{2} A_A{}^2\right\} + \sum_{A=3}^{4}\left(a_A{}^\dagger a_A + a_A a_A{}^\dagger\right)\right]. \tag{3.31}$$

The quark coordinates and momenta are given by (3.21) and (3.25), respectively. With the solutions (3.29), we can write for the two systems:

<u>Lepton quarks</u>

$$Q_{IA} = \left[A_{IA} + (2N)^{-1/2} A_1\right] s + \left[B_{IA} + (2N)^{-1/2} B_1\right]$$
$$\pm \frac{1}{2\sqrt{NM}}\left[a_3{}^\dagger \exp(i\omega s) + a_3 \exp(-i\omega s)\right],$$

$$P_{IA} = (M/\omega)\left[A_{IA} + (2N)^{-1/2} A_1\right]$$
$$\pm \frac{i}{2}\sqrt{\frac{M}{N}}\left[a_3{}^\dagger \exp(i\omega s) - a_3 \exp(-i\omega s)\right], \quad A = 1,\ 3; \tag{3.32}$$

<u>Hadron quarks</u>

$$Q_{IA} = \left[A_{IA} + (2N)^{-1/2} A_2\right] s + \left[B_{IA} + (2N)^{-1/2} B_2\right]$$
$$\pm \frac{1}{2\sqrt{NM}}\left[a_4{}^\dagger \exp(i\omega s) + a_4 \exp(-i\omega s)\right],$$

$$P_{IA} = (M/\omega)\left[A_{IA} + (2N)^{-1/2} A_2\right]$$
$$\pm \frac{i}{2}\sqrt{\frac{M}{N}}\left[a_4{}^\dagger \exp(i\omega s) - a_4 \exp(-i\omega s)\right], \quad A = 2,\ 4. \tag{3.33}$$

Examining these solutions, it is apparent that pairs of quarks in either system can be paired up by boundary or initial conditions into composite particles. For example, setting $A_{I2}=A_{I4}$ and $B_{I2}=B_{I4}$ yields a system of hadron composites described by

$$Q_I^H \equiv \tfrac{1}{2}(Q_{I2} + Q_{I4}) = \left[A_{IA} + (2N)^{-1/2} A_2\right] s + (B_{I2} + B_2),$$
$$P_I^H \equiv (P_{I2} + P_{I4}) = 2(M/\omega)\left[A_{IA} + (2N)^{-1/2} A_2\right]. \tag{3.34}$$

The internal state of these hadron composites are all described by the same function

$$q_I^H \equiv \tfrac{1}{2}(Q_{I2} - Q_{I4}) = \tfrac{1}{2}(NM)^{-1/2}\left[a_4{}^\dagger \exp(i\omega s) + a_4 \exp(-i\omega s)\right]. \tag{3.35}$$

In similar fashion, we can compose lepton composites $Q_I^L$, $P_I^L$, and $q_I^L$.

### C. Natural Boundary Conditions and the Lagrangian Constraints

As we indicated in Sec. II, application of the variational principle implies preferred reference frames, and we have assumed these frames are those for which the coordinates are the position and velocity vectors of "real" particles. Application of the variational principle to the Lagrangian (3.1) yields a set of natural boundary conditions (n.b.c.), which can be re-expressed in terms of the quark momenta as

$$P_{IA}{}^2(s) \ \text{-}\ 0 \quad \text{as } s\ 6\ \text{"}\ 4, \qquad (3.36)$$

where $P_{IA}(s)$ is given by (3.32) or (3.33).

In going over to the quantized model, one might expect to write the n.b.c. as

$$P_{IA}{}^2(s)\left|\Psi\right\rangle \ \text{-}\ 0 \quad \text{as } s\ 6\ \text{"}\ 4\ . \qquad (3.37)$$

However, the commutation relations (3.28) lead to inconsistencies. Thus, in analogy to oscillator constraint relations in quantum field theory, we apply the n.b.c. in the following form:

$$\left[A_{IA} + (NM)^{1/2} A_1\right]^2 \left|\Psi\right\rangle = -\left(\omega^2/4NM\right)\left(a_3{}^\dagger a_3 + a_3\, a_3{}^\dagger\right)\left|\Psi\right\rangle,$$

$$a_A{}^2\left|\Psi\right\rangle = 0, \qquad \left[A_{IA} + (2N)^{-1/2} A_1\right] a_3 \left|\Psi\right\rangle = 0, \quad A = 1,\ 3;$$

$$\left[A_{IA} + (NM)^{1/2} A_2\right]^2 \left|\Psi\right\rangle = -\left(\omega^2/4NM\right)\left(a_4{}^\dagger a_4 + a_4\, a_4{}^\dagger\right)\left|\Psi\right\rangle,$$

$$a_A{}^2\left|\Psi\right\rangle = 0, \qquad \left[A_{IA} + (2N)^{-1/2} A_2\right] a_4 \left|\Psi\right\rangle = 0, \quad A = 2,\ 4. \qquad (3.38)$$

The equations in (3.38) imply $\left\langle\Psi\right|\ a_A{}^{\dagger 2} = 0$, etc., so that the expectation value of $P_{IA}{}^2(s)$ is zero as $s\ 6\ \text{"}\ 4$.

The eigenstates of H can be labeled by the eigenvalues of $A_{IA}$, $A_A$, and $n_A$ / $- a_A{}^\dagger a_A$. Then, for the composite solutions such as (3.34), the n.b.c. contain the mass-shell conditions

$$P_I{}^2 = -\frac{M}{N}\left(a_A{}^\dagger a_A + a_A a_A{}^\dagger\right) = (2M/N)\left(n_A + 2\right). \qquad (3.39)$$

The n.b.c. also imply that for a positive mass-squared composite particle in its rest frame

$$a_{A0}\left|\Psi\right\rangle = a_{A0}\left|A_{IA}, A_A, n_A\right\rangle = 0. \qquad (3.40)$$

In other words, there are no time oscillations in the rest frame of positive mass-squared particles.

Using the relation $\sum_{I=1}^{N} A_{IA} = 0$, we can recast the Hamiltonian (3.31) as

$$H=\frac{\omega}{2}\left[\frac{M}{\omega^2}\sum_{I=1}^{N}\sum_{A=1}^{4}\left[A_{IA}+(2N)^{-1/2}A_A\right]^2+\sum_{A=3}^{4}\left(a_A{}^{\dagger}a_A+a_Aa_A{}^{\dagger}\right)\right]. \qquad (3.41)$$

Substitution of the mass-shell conditions in (3.38) into (3.41) implies that H=0 can be satisfied only if

$$\sum_{A=3}^{4}\left(a_A{}^{\dagger}a_A+a_A\,a_A{}^{\dagger}\right)=-2\sum_{A=3}^{4}\left(n_A+2\right)=0. \qquad (3.42)$$

Thus, for a system of lepton and hadron composites, one or both types of composites must have negative squared mass.

Consider now the constraint problem. Rephrase the $\Phi_i$ in terms of quantized operators:

$$\Phi_1 \equiv \mathbf{P}^2+M^2\mathbf{Q}\mathbf{G}^2\mathbf{Q},$$

$$\Phi_2 \equiv \frac{1}{2}\left(\mathbf{PGQ}+\mathbf{QGP}\right),$$

$$\Phi_3 \equiv \left(\mathbf{PGP}-\mathbf{QGQ}\right). \qquad (3.43)$$

Substitution of the solutions (3.21), (3.25), and (3.29) into the last two relations yields

$$\Phi_2=\sum_{A=3}^{4}\left(p_AW_A+W_Ap_A\right)=i\sum_{A=3}^{4}\left[a_A{}^{\dagger 2}\exp(2i\omega s)-a_A{}^2\exp(-2i\omega s)\right], \qquad (3.44)$$

$$\Phi_3=\sum_{A=3}^{4}\left(p_A{}^2-M^2W_A{}^2\right)=-M\sum_{A=3}^{4}\left[a_A^{\dagger 2}\exp(2i\omega s)+a_A^2\exp(-2i\omega s)\right]. \qquad (3.45)$$

Thus, $\Phi_2 \approx 0$ and $\Phi_3 \approx 0$ are satisfied by the n.b.c. (note that the converse is not true, i.e., the constraint relations above do not imply the n.b.c.).

## IV. HARMONIC OSCILLATOR QUARK MODEL WITH SPONTANEOUS BROKEN SYMMETRY

### A. Breaking the Lepton-Hadron Symmetry

The lepton-hadron symmetric Lagrangian (3.1) must be modified in order to obtain a description of interacting composite particles. This is accomplished here, not by the addition of an interaction term, but by adopting an analog to the familiar prescription of adding a positive infinitesimal imaginary part to the mass in particle scattering problems. We shall find that as a result, the lepton-hadron symmetry is broken and we are led to a more realistic model.

We begin by replacing the harmonic oscillator potential of (3.6) by

$$V = -\frac{M^2(1+2i\varepsilon)}{2N}\sum_{I=1}^{N}\sum_{J=1}^{N}\sum_{A=1}^{4}\sum_{B=1}^{4}\left(\mathbf{h}^{\dagger}\mathbf{h}\right)_{AB}\left[x_{IA}(s)-x_{JB}(s)\right]^2. \tag{4.1}$$

The new coupling matrix $\mathbf{h}$ is defined as

$$\mathbf{h} \equiv \mathbf{g} + i\varepsilon\mathbf{d}, \tag{4.2}$$

where $\varepsilon$ is a positive infinitesimal parameter, and $\mathbf{g}$ and $\mathbf{d}$ are defined as

$$\mathbf{g} = \frac{1}{2}\begin{pmatrix} 1 & 0 & -1 & 0 \\ 0 & 1 & 0 & -1 \\ -1 & 0 & 1 & 0 \\ 0 & -1 & 0 & 1 \end{pmatrix} = \mathbf{g}^2, \tag{3.4}$$

$$\mathbf{d} \equiv \frac{1}{4}\begin{pmatrix} 1 & 1 & 1 & 1 \\ 1 & 1 & 1 & 1 \\ 1 & 1 & 1 & 1 \\ 1 & 1 & 1 & 1 \end{pmatrix} = \mathbf{d}^2. \tag{4.3}$$

Since $\mathbf{g}\cdot\mathbf{d} = \mathbf{d}\cdot\mathbf{g} = 0$, we have

$$\mathbf{h}^{\dagger}\mathbf{h} = \mathbf{g}^2 + \varepsilon^2\mathbf{d}^2. \tag{4.4}$$

Note that if we define a real matrix $\mathbf{h}' \equiv \mathbf{g} + \varepsilon\mathbf{d}$, then $\mathbf{h}'^2 = \mathbf{h}^{\dagger}\mathbf{h}$.

Thus we have modified the symmetric Lagrangian (3.1) by the replacements $\mathbf{g} \to \mathbf{h}$ and $M \to (1+i\varepsilon)M$. The limit $\varepsilon \to 0$ is to be taken after calculation of a matrix element. Note that V retains a symmetry analogous to that of Lagrangian (3.1), i.e., it is symmetric under the exchange $x_{I1} \leftrightarrow x_{I2}$ and $x_{I3} \leftrightarrow x_{I4}$. Because the potential exhibits this basic symmetry, we shall regard the coordinates $x_{IA}(s)$ as a set of *fundamental* coordinates, but not the coordinates of the quarks which constitute measurable composite particles (see the discussion in Sec. II).

The matrix $\mathbf{h}^{\dagger}\mathbf{h}$ (or the matrix $\mathbf{h}'$) is diagonalized, not by the matrix $\delta$ in the hadron-lepton symmetric model, but by the matrix defined below:

$$\rho \equiv \frac{1}{2}\begin{bmatrix} 1 & 1 & 1 & 1 \\ 1 & -1 & 1 & -1 \\ 1 & 1 & -1 & -1 \\ 1 & -1 & 1 & -1 \end{bmatrix} = \rho^{-1}. \tag{3.19}$$

Rather than solve for the coordinates $x_{IA}$, however, we shall make a transformation to what will be defined as the quark coordinates $Q_{IA}$:

$$\mathbf{Q} \equiv (\xi \otimes \mathbf{1})\mathbf{x}, \tag{4.5}$$

where

$$\xi \equiv \frac{1}{\sqrt{2}}\begin{bmatrix} 1 & 1 & 0 & 0 \\ 1 & -1 & 0 & 0 \\ 0 & 0 & 1 & 1 \\ 0 & 0 & 1 & -1 \end{bmatrix} = \xi^{-1}. \tag{4.6}$$

Note that $\xi \cdot \rho = \delta$ and $\xi^{-1}\mathbf{g}\xi = \mathbf{g}$.

The potential can now be put into the form

$$V = (M\mathbf{H}\mathbf{Q})^2 \tag{4.7}$$

where

$$\mathbf{H} \equiv (\mathbf{g} - i\varepsilon\mathbf{f}) \otimes \mathsf{N} + i\varepsilon\, \mathbf{1} \otimes \mathbf{1}, \tag{4.8}$$

and **f** is defined as

$$\mathbf{f} \equiv \xi^{-1}\mathbf{d}\xi = \frac{1}{2}\begin{bmatrix} 1 & 0 & 1 & 0 \\ 0 & 0 & 0 & 0 \\ 1 & 0 & 1 & 0 \\ 0 & 0 & 0 & 0 \end{bmatrix} = \mathbf{f}^2 . \tag{4.9}$$

The matrix $\mathsf{N}$ is given by (3.4). Note that $\mathbf{f} \cdot \mathbf{g} = \mathbf{g} \cdot \mathbf{f} = 0$.

The "stringlike" Lagrangian to which the variational principle will be applied becomes

$$L = -7M^2\left[-(\mathbf{H}\mathbf{Q})^2\, \dot{\mathbf{Q}}^2 + (\dot{\mathbf{Q}}\,\mathbf{H}\mathbf{Q})^2\right]^{1/2}. \tag{4.10}$$

This Lagrangian does not display a symmetry under $Q_{I1} \leftrightarrow Q_{I2}$, $Q_{I3} \leftrightarrow Q_{I4}$. The momentum conjugate to **Q** is

$$\mathbf{P} = \frac{\partial L}{\partial \dot{\mathbf{Q}}} = \frac{M^2}{L}\left[ -\left(\mathbf{HQ}\right)^2 \dot{\mathbf{Q}} + \left(\dot{\mathbf{Q}}\mathbf{HQ}\right)\mathbf{HQ} \right]. \tag{4.11}$$

The nonstandard Lagrangians (3.1) and (4.10) do not yield Hamiltonians, but only constraints. However, contrary to the parallel problems in gravity[11] and string theory,[24,13,19,20] this does not complicate the quantization procedure for either Lagrangian. To construct a Hamiltonian corresponding to (4.10), we turn again to Dirac's formulation of the constraint problem.[29] Two primary constraints result from L:

$$\Phi_1 \equiv \mathbf{P}^2 + M^2 \left(\mathbf{HQ}\right)^2 \approx 0, \tag{4.12}$$

and

$$\Phi_2 \equiv \mathbf{PHQ} \approx 0. \tag{4.13}$$

Begin by taking the Hamiltonian to be $H = v_1 \Phi_1 + v_2 \Phi_2$. Consistency requires $\{\Phi_i, H\} \approx 0,\ i = 1,2,$ or

$$\{\Phi_1, \Phi_2\} = 2\Phi_3 \approx 0, \tag{4.14}$$

where we define

$$\Phi_3 \equiv \mathbf{PGP} - M^2 \mathbf{QGQ}. \tag{4.15}$$

Now modify the Hamiltonian to read

$$H = \sum_{i=1}^{3} v_i \Phi_i. \tag{4.16}$$

Once more, consistency requires $\{\Phi_i, H\} \approx 0,\ i = 1,2,3$. This yields a set of equations (again, to first order)

$$\left(4M^2 \mathbf{PGQ}\right) v_3 \approx 0,$$

$$\left(\mathbf{PGP} + M^2 \mathbf{QGQ}\right) v_3 \approx 0,$$

$$\left(4M^2 \mathbf{PGQ}\right) v_1 - 2\left(\mathbf{PGP} + M^2 \mathbf{QGQ}\right) v_2 \approx 0. \tag{4.17}$$

Thus, for a nontrivial solution for the $v_i$'s, we must add at least one more constraint. Further examination of the constraint procedure shows that the least restrictive solution is obtained by setting $v_2=v_3=v_4=0$ and defining a secondary constraint (to first order)

$$\Phi_4 \equiv \mathbf{PGP} \approx 0. \tag{4.18}$$

Finally, choosing a gauge $v_1=\omega/2M$, we take the Dirac Hamiltonian to be

$$H=(\omega/2M)\left[\mathbf{P}^2+M^2(\mathbf{HQ})^2\right]. \tag{4.19}$$

The equations of motion

$$\dot{\mathbf{P}}=-\frac{\partial H}{\partial \mathbf{Q}}, \qquad \dot{\mathbf{Q}}=\frac{\partial H}{\partial \mathbf{P}}, \tag{4.20}$$

yield

$$\ddot{\mathbf{Q}}=-\omega^2\mathbf{H}^2\mathbf{Q}, \qquad \mathbf{P}=(M/\omega)\dot{\mathbf{Q}}. \tag{4.21}$$

Following Sec. III, define a transformation of coordinates

$$Q_{IA}=y_{IA}+N^{-1/2}(\delta\mathbf{W})_A, \qquad \sum_{I=1}^{N} y_{IA}=0. \tag{4.22}$$

The matrix $\delta$ appeared in Sec. III, and is given by

$$\delta\equiv\frac{1}{\sqrt{2}}\begin{pmatrix}1&0&1&0\\0&1&0&1\\1&0&-1&0\\0&1&0&-1\end{pmatrix}=\delta^{-1}. \tag{3.20}$$

The momentum relation in (4.21) yields

$$P_{IA}=(M/\omega)\dot{Q}_{IA}=(M/\omega)\left[\dot{y}_{IA}+N^{-1/2}(\delta\dot{\mathbf{W}})_A\right]. \tag{4.23}$$

This suggests defining momenta $p_{IA}$ and $p_A$ by

$$p_{IA}\equiv(M/\omega)\dot{y}_{IA}, \qquad \sum_{I=1}^{N} p_{IA}=0; \qquad p_A\equiv(M/\omega)\dot{W}_A. \tag{4.24}$$

Then,

$$P_{IA}=p_{IA}+N^{-1/2}(\delta\mathbf{p})_A. \tag{4.25}$$

In terms of the transformed coordinates, the Hamiltonian becomes

$$H=\frac{\omega}{2M}\Big\{\sum_{I=1}^{N}\sum_{A=1}^{4}\left[p_{IA}{}^2-\varepsilon^2M^2y_{IA}{}^2\right]+p_1{}^2+\left[p_2{}^2-\varepsilon^2M^2W_2{}^2\right] + \sum_{A=3}^{4}\left[p_A{}^2+(1+i\varepsilon)^2M^2W_A{}^2\right]\Big\}. \tag{4.26}$$

The momenta $p_{IA}$ and $p_A$ are conjugate to $y_{IA}$ and $W_A$, respectively.

Using the Hamiltonian above or the equations of motion (4.21), we obtain

$$\ddot{y}_{IA} = \varepsilon^2 \omega^2 y_{IA}, \qquad \ddot{W}_1 = 0; \qquad \ddot{W}_2 = \varepsilon^2 \omega^2 W_2;$$

$$\ddot{W}_A = -\omega^2 (1 + i\varepsilon)^2 W_A, \qquad A = 3,\ 4. \tag{4.27}$$

## B. The quantized solutions: Quarks as strings

We shall quantize by imposing the quantum conditions

$$[y_{IA\mu}, p_{IA\nu}] = -i\, g_{\mu\nu}, \quad I = 1,\ 2,\ 3,\ \ldots,\ N-1; \qquad [W_{A\mu}, p_{A\nu}] = -i\, g_{\mu\nu}. \tag{4.28}$$

Working in the Heisenberg Picture, we use $\dot{O} = -i\,[O, H]$ to obtain the equations of motion which are the operator forms of (4.27). Express the solutions as

$$y_{IA} = (1/\varepsilon)(2M)^{-1/2}\left[a_{IA} \exp(\varepsilon\omega s) + b_{IA} \exp(-\varepsilon\omega s)\right],$$

$$p_{IA} = (M/2)^{1/2}\left[a_{IA} \exp(\varepsilon\omega s) - b_{IA} \exp(-\varepsilon\omega s)\right];$$

$$W_1 = A_1 s + B_1, \qquad p_1 = (M/\omega) A_1;$$

$$W_2 = (1/\varepsilon)(2M)^{-1/2}\left[a_2 \exp(\varepsilon\omega s) + b_2 \exp(-\varepsilon\omega s)\right],$$

$$p_2 = (M/2)^{1/2}\left[a_2 \exp(\varepsilon\omega s) - b_2 \exp(-\varepsilon\omega s)\right];$$

$$W_A = (2M)^{-1/2}\left[a_A^{\dagger} \exp(i(1+i\varepsilon)\omega s) + a_A \exp(-i(1+i\varepsilon)\omega s)\right],$$

$$p_A = i(M/2)^{1/2}(1+i\varepsilon)\left[a_A^{\dagger} \exp(i(1+i\varepsilon)\omega s) - a_A \exp(-i(1+i\varepsilon)\omega s)\right], \quad A = 3,\ 4. \tag{4.29}$$

The choice of normalization implies that $y_{IA}$ and $W_2$ become singular as $\varepsilon \to 0$. Thus, a decomposition of $H=H_0+H_I$ has no meaning, since $\varepsilon y_{IA}$ and $\varepsilon W_2$ remain finite. In quantizing, these singularities are not a problem if we work in the momentum representation and not allow the limit $\varepsilon \to 0$ to be taken until after calculation of a matrix element. The physical interpretation of the position space as $\varepsilon \to 0$ will be given later when composite scattering is considered.

The commutation relations (4.28) imply

$$[a_{IA\mu}, b_{IA\nu}] = i\varepsilon\, g_{\mu\nu}; \quad [a_{2\mu}, b_{2\nu}] = i\,\varepsilon\, g_{\mu\nu};$$

$$\left[a_{A\mu}, a_{A\nu}{}^{\dagger}\right] = -(1+i\varepsilon)^{-1} g_{\mu\nu}. \tag{4.30}$$

The Hamiltonian becomes

$$H = \frac{\omega}{2}\left[\left(M/\omega^2\right) A_1{}^2 - \sum_{I=1}^{N}\sum_{A=1}^{4}\left(a_{IA}b_{IA} + b_{IA}a_{IA}\right) - \left(a_2 b_2 + b_2 a_2\right) + (1+i\varepsilon)^2 \sum_{A=3}^{4}\left(a_A{}^{\dagger} a + a_A a_A{}^{\dagger}\right)\right]. \tag{4.31}$$

If, at this point, we were to impose the constraint H=0, we would obtain from the above an expression for $\omega$ depending on the various eigenvalues $a_{IA}$, $b_{IA}$, etc. This would imply that each final state quark carries a "memory" of the process which created them. Furthermore, the parameter $\varepsilon$ introduces an infinitesimal repulsive harmonic-oscillator potential. To avoid these problems, we shall choose the gauge parameter $\omega$ as follows:

$$\omega = \omega_0 / \varepsilon. \tag{4.32}$$

Later, we shall see that this is equivalent to an absence of a background frame of reference for the fundamental coordinates $x_{IA}$. In view of this gauge choice, we can replace, to *first* order, commutation relations in (4.28) by the following:

$$\left[a_{IA\mu}, b_{IA\nu}\right] = 0; \qquad \left[a_{2\mu}, b_{2\nu}\right] = 0. \tag{4.33}$$

Now the exponential terms in the earlier solutions become

$$\exp\left(-\varepsilon\omega s\right) = \exp\left[-\sqrt{\varepsilon}\,\omega_0 s\right], \qquad \exp\left(-i(1+i\varepsilon)\omega s\right) = \exp\left[-i(1+i\varepsilon)\omega_0 s/\sqrt{\varepsilon}\right]. \tag{4.34}$$

The second of these relations shows that, as $\varepsilon$ tends to zero, the oscillation frequency of the quarks tends to infinity. That is, the quarks approach becoming closed strings, mapping out tubes in space time. It is important to note that these "strings" *are not physically extended objects. In other words, they are not the strings of string theories.* As a result, the problem of an infinite number of constraints (nonclosure of the constraint algebra) is avoided, and the theory is formulated consistently in four dimensional space time.

The choice of gauge (4.32) allows us to drop the term $p_1{}^2 = \left(M^2/\omega^2\right) A_1{}^2$ in (4.26) and (4.31). To first order, we now can write the Hamiltonian as

$$H = \frac{\omega}{2}\left[-\sum_{I=1}^{N}\sum_{A=1}^{4}\left(a_{IA}b_{IA} + b_{IA}a_{IA}\right) - \left(a_2 b_2 + b_2 a_2\right) + \sum_{A=3}^{4}\left(a_A{}^{\dagger} a + a_A a_A{}^{\dagger}\right)\right]. \tag{4.35}$$

This is a significant result, for if the term $p_1^2$ had to be retained, the frequency would depend on the constants of integration.

### C. The natural boundary conditions (n.b.c.)

The variational principle of Sec. II applied to the Lagrangian (4.10) yields the classical natural boundary conditions (n.b.c.)

$$P_{IA}^2(s) \to 0 \text{ as } s \to \pm\infty, \tag{4.36}$$

with the momentum $P_{IA}(s)$ given by (4.25). Before applying the n.b.c., consider again the solutions (4.29). The oscillatory variables $W_A$ and $p_A$, A=3, 4, are complex numbers. We shall therefore implicitly understand them to be replaced with the variables $(1/2)(W_A+W_A^\dagger)$ and $(1/2)(p_A+p_A^\dagger)$ , and write the solutions to first order as

$$W_A = (2M)^{-1/2}\cosh\left(\sqrt{\varepsilon}\,\omega_0 s\right)\left[a_A^\dagger \exp\left(i\omega_0 s/\sqrt{\varepsilon}\right) + a_A \exp\left(-i\omega_0 s/\sqrt{\varepsilon}\right)\right],$$

$$p_A = i\,(M/2)^{1/2}\cosh\left(\sqrt{\varepsilon}\,\omega_0 s\right)\left[a_A^\dagger \exp\left(i\omega_0 s/\sqrt{\varepsilon}\right) - a_A \exp\left(-i\omega_0 s/\sqrt{\varepsilon}\right)\right]. \tag{4.37}$$

For $\varepsilon \to 0$, substitution of (4.37) into the commutation relation (4.28) now yields

$$\left[a_{A\mu}, a_{A\nu}^\dagger\right] = -\frac{1}{\cosh\sqrt{\varepsilon}\,\omega_0 s}\, g_{\mu\nu}. \tag{4.38}$$

This would appear to present a problem, for $a_{IA\mu}$ and $a^\dagger{}_{IA\nu}$ are supposed to be constant operators. However, we have not, up till now, discussed the order of limit taking for s and $\varepsilon$. We shall adopt the following prescription, while not mathematically rigorous, gives us a set of rules with which to work, as long as they are applied consistently. We shall always assume $\sqrt{\varepsilon}$ is finite as s approaches the asymptotic regions $\pm\infty$. The matrix elements are then calculated, and finally we allow $\varepsilon$ to go to zero. Thus, as $s \to \pm\infty$ in (4.38), we have

$$\lim_{s\to\pm\infty}\left[a_{A\mu}, a_{A\nu}^\dagger\right] = 0, \text{ for } \sqrt{\varepsilon} \to 0. \tag{4.39}$$

In other words, for $\sqrt{\varepsilon} \to 0$, and in the *asymptotic regions*, we treat $a_{IA\mu}$ and $a^\dagger{}_{IA\nu}$ as commuting operators.

With the replacements (4.37), the Hamiltonian is only approximately constant in s, of course,

until $\varepsilon \to 0$. To first order, H becomes

$$H = \frac{\omega}{2}\left[-\sum_{I=1}^{N}\sum_{A=1}^{4}\left(a_{IA}b_{IA} + b_{IA}a_{IA}\right) - \left(a_2 b_2 + b_2 a_2\right) + \cosh^2\left(\sqrt{\varepsilon}\,\omega_0 s\right)\sum_{A=3}^{4}\left(a_A{}^{\dagger} a + a_A a_A{}^{\dagger}\right)\right]. \quad (4.40)$$

We will use this form of H in Sec. VI to go from the Heisenberg Picture to the Schrödinger Picture for application of electroweak perturbation techniques.

We need now to discuss the labeling of states. In the limit $\varepsilon \to 0$, we can take the set of commuting variables $a_{IA}$, $b_{IA}$, $a_2$, $b_2$, and $n_A \equiv -a_A{}^{\dagger} a_A$, and write

$$\left|\Psi\right\rangle \equiv \left|a_{IA}, b_{IA}, a_2, b_2, n_A\right\rangle. \quad (4.41)$$

We shall retain this labeling for the asymptotic states even in the case $\varepsilon \neq 0$, since to first order $a_{IA}$ and $b_{IA}$ still commute.

Returning now to the n.b.c., we follow the procedure of Sec.III and write the quantized conditions as follows:

Lepton quarks (A=1,3)

$$a_{IA}{}^2\left|\Psi\right\rangle = b_{IA}{}^2\left|\Psi\right\rangle = -\left(1/8N\right)\left(a_3{}^{\dagger}a_3 + a_3 a_3{}^{\dagger}\right)\left|\Psi\right\rangle = \frac{1}{4N}\left(n_3 + 2\right)\left|\Psi\right\rangle,$$

$$a_3{}^2\left|\Psi\right\rangle = 0, \qquad a_{IA}\, a_3\left|\Psi\right\rangle = 0, \qquad b_{IA}\, a_3\left|\Psi\right\rangle = 0. \quad (4.42)$$

Hadron quarks (A=2,4)

$$\left[a_{IA} + \left(2N\right)^{-1/2} a_2\right]^2\left|\Psi\right\rangle = \left[b_{IA} + \left(2N\right)^{-1/2} b_2\right]^2\left|\Psi\right\rangle = \frac{1}{4N}\left(n_4 + 2\right)\left|\Psi\right\rangle,$$

$$a_4{}^2\left|\Psi\right\rangle = 0, \quad \left[a_{IA} + \left(2N\right)^{-1/2} a_2\right] a_4\left|\Psi\right\rangle = 0, \quad \left[b_{IA} + \left(2N\right)^{-1/2} b_2\right] a_4\left|\Psi\right\rangle = 0. \quad (4.43)$$

In deriving (4.42) and (4.43), we have made use of the relations $\sum_{I=1}^{N} a_{IA} = 0$, $\sum_{I=1}^{N} b_{IA} = 0$.

### D. Satisfaction of the Dirac Constraints by the n.b.c.

We write the quantized versions of the constraints as

$$\Phi_1 \equiv \mathbf{P}^2 + M^2\mathbf{QH}^2\mathbf{Q} \approx 0,$$

$$\Phi_2 \equiv \frac{1}{2}\left(\mathbf{PHQ} + \mathbf{QHP}\right) \approx 0,$$

$$\Phi_3 \equiv \mathbf{PGP} - M^2\mathbf{QGQ} \approx 0,$$

$$\Phi_4 \equiv \frac{1}{2}\left(\mathbf{PGP} + \mathbf{QGP}\right) \approx 0. \tag{4.44}$$

It is not difficult to demonstrate that the solutions (4.29), along with the n.b.c., ensure the vanishing of $\Phi_2$, $\Phi_3, \Phi_4$. Thus, aside from the condition $H \approx 0$, all constraints are automatically satisfied by the application of the n.b.c. (the converse is not true).

One might ask, then, why not keep the simpler Lagrangian of the form $L = \sqrt{VT}$ ? One answer lies in the possibility of generalizing the Lagrangian to describe other forces. For example, write

$$L = -\left\{-[\mathbf{F}(\mathbf{Q})]^2\dot{\mathbf{Q}}^2 + [\mathbf{F}(Q)\cdot\dot{\mathbf{Q}}]^2\right\}^{1/2}, \tag{4.45}$$

where **F(Q)** takes on different forms depending on the conditions of the domain of space-time being considered. In Sec.VI, in a domain where the electroweak forces dominate, L is assumed to take the form

$$L = -\left\{-[\mathbf{A}(\mathbf{Q})]^2\,\dot{\mathbf{Q}}^2 + [\mathbf{A}(Q)\cdot\dot{\mathbf{Q}}]^2\right\}^{1/2}, \tag{4.46}$$

where **A(Q)** is the electroweak potential. This Lagrangian is used to calculate spinless lepton-lepton electroweak scattering amplitudes.

### E. Quark-Antiquark assumption

The variables $W_A$, which are shared by different quarks, can be considered associated with the internal symmetries of the composite leptons and hadrons. In earlier work,[26] the coefficients of $W_1$ and $W_2$ were denoted as the P and X numbers respectively, and the coefficients of $W_3$ and $W_4$ as the L and B numbers, respectively. The leptons (hadrons) are composed are composed of pairs of quarks of the same P (X) number and opposite L (B) number.

There has been nothing in the formulation thus far that implies the necessity for what could be termed antiquarks. Feynman[30] gives a nice demonstration of why quantum mechanics and the relativistic restriction of particle velocities less than c require the existence of antiparticles. However, the quark velocities of this model are not so restricted, although the velocities of the observable composites are.

In order to produce composite particles which are accompanied by corresponding antiparticles, we shall introduce a definition of an antiquark. Recall that an electron going backward in time can be re-interpreted as a positron of opposite spin going forward in time. Now consider the variables $Q_{IA}(-s)$ which describe quarks going the reverse sense in the variable s:

Lepton quarks

$$\overline{Q}_{IA}(s) \equiv Q_{IA}(-s) = \overline{y}_{IA}(s) + \frac{1}{\sqrt{2N}} W_1(-s) + (-1)^{\frac{A-1}{2}} \frac{1}{\sqrt{2N}} W_3(-s). \tag{4.47}$$

Hadron quarks

$$\overline{Q}_{IA}(s) \equiv Q_{IA}(-s)$$
$$= \overline{y}_{IA}(s) + \frac{1}{2\varepsilon\sqrt{NM}} \left[ b_2 \exp\left(i\sqrt{\varepsilon}\,\omega_0 s\right) + a_2 \exp\left(-i\sqrt{\varepsilon}\,\omega_0 s\right) \right] + (-1)^{\frac{A-2}{2}} W_4(-s). \tag{4.48}$$

In both (4.47) and (4.48), we have set $\overline{y}_{IA}(s) \equiv y_{IA}(-s)$. In the asymptotic regions of s, $W_1(-s)$ plays no role. Thus, lepton quarks going "backwards" in s can be reinterpreted as quarks with opposite internal angular momentum going "forward" in s. In the case of the hadron quarks, consider solutions for which

$$a_2 = b_2. \tag{4.49}$$

For these solutions, the hadron quarks going "backwards" in s can be reinterpreted the same way. We shall define such solutions as antiquarks and postulate that the quarks that constitute initial-state observable hadrons correspond to solutions satisfying (4.49). Therefore, the hadrons can be thought of as composed of quark and antiquark.

In the next section, it will be seen that $W_1$ plays no role in the "strong" scattering of leptons. The condition $a_2 = b_2$ is necessary for the conservation of four-momentum in hadron scattering. It also leads to final state hadrons similarly composed of quark-antiquark pairs.

It is important to note that hadron-hadron four-momentum conservation is not built in by a basic symmetry of the Lagrangian, but is closely tied to the particle-antiparticle requirements of quantum mechanics and special relativity.

## F. The absence of a background frame of reference

At the beginning of this section, we assumed the $x_{IA}$ to be the fundamental in the sense that the Lagrangian in these coordinates display a basic symmetry $x_{I1} \leftrightarrow x_{I2}$, $x_{I3} \leftrightarrow x_{I4}$. The solutions for $x_{IA}$ are easily obtained from (4.5) and (4.22). They take the form

$$x_{IA} = \langle \xi \mathbf{y}_I \rangle_A + N^{-1/2} \langle \rho \mathbf{W} \rangle_A . \qquad (4.50)$$

The momenta, which we will define as $p_{IA}$, are similarly obtained:

$$p_{IA} = \langle \xi \mathbf{p}_I \rangle_A + N^{-1/2} \langle \rho\, \mathbf{p} \rangle_A . \qquad (4.51)$$

From (4.50), we obtain a center-of-mass vector

$$X \equiv \frac{1}{4N} \sum_{I=1}^{N} \sum_{A=1}^{4} x_{IA} = \frac{1}{2\sqrt{N}} W_1 , \qquad (4.52)$$

while (4.51) yields the total momentum

$$P \equiv \sum_{I=1}^{N} \sum_{A=1}^{4} p_{IA} = 2\sqrt{N}\, p_1 . \qquad (4.53)$$

We see from the solution in (4.29) for $W_1$ that the center-of-mass vector plays no role in the formation of composite particles in the asymptotic regions $s \to \pm\infty$. From the solution for $p_1$ in (4.29) and the gauge choice (4.32), it follows that the total momentum P vanishes as $\varepsilon \to 0$. In similar manner, the total angular momentum vanishes also. Thus, there is no background frame of reference. All composite variables are expressed in terms of relational coordinates.

## G. Representations and completeness relations

The natural boundary conditions (n.b.c.) imply that the solutions of the *dynamical* system are a subset of the simultaneous eigenstates of the commuting operators $a_{IA}$, $b_{IA}$, $a_2$, $b_2$, and $n_A$=-$a_A^{\dagger} a_A$. Dirac[21] defines a complete set of eigenvalues of a real dynamical variable as just the possible results of measurements of that dynamical variable. Let us modify that definition: The complete set of eigenvalues of a set of commuting real dynamical variables is determined by the possible solutions of the dynamical system. To see what that implies for the present model, let us first look a the simple case of a lepton quark formed as $s \to +\infty$. The results are easily generalized to hadron quarks. Label the eigenstates of $a_{IA}$ and $n_3$ which satisfy n.b.c. conditions (4.42) as $|a_{IA}, n_3\rangle$.

That is, these states satisfy

$$a_3{}^2\,|a_{IA},n_3\rangle = 0, \text{ and } a_{IA}\cdot a_3\,|a_{IA},n_3\rangle = 0. \tag{4.54}$$

We must, of course, also include the mass shell condition in (4.42). We do this by defining the completeness relation

$$1 = \sum_{n_3=-\infty}^{\infty}\int_{-\infty}^{\infty} d^4 a_{IA}\,|a_{IA},\, n_3\rangle\,\delta\left[a_{IA}{}^2 - \mu_{n_3}{}^2\right]\langle a_{IA},\, n_3|, \tag{4.55}$$

with

$$\mu_{n_3}{}^2 \equiv \frac{1}{4N}\left(n_3 + 2\right). \tag{4.56}$$

Examination of the expression for $p_{IA}$ in (4.29) indicates that, as $s \to +\infty$, $|a_{IA},\, n_3\rangle$ is an eigenstate of $p_{IA}$

Until now we have paid little attention to the position vector $y_{IA}$. As we shall see in the next section, such a representation for hadron-hadron or lepton-lepton scattering plays no role in the calculation of the strong-interaction scattering amplitudes. However, in Sec. VI, we will look at perturbation theory applied to electroweak scattering. There, we will use the asymptotic harmonic-oscillator states as the unperturbed states and introduce the position representation as intermediate states. From (4.29), we see that $y_{IA}$ is proportional to $a_{IA}/\varepsilon$, and thus is singular as $\varepsilon \to 0$.

However, since $\varepsilon$ is kept finite until a matrix element is calculated, we can define a position representation when the states are summed over as intermediate states in the matrix element.

We shall label the position states as $s \to +\infty$ as $|a_{IA}/\varepsilon, n_3\rangle$, and assume the completeness relation

$$1 = \sum_{n_3=-\infty}^{\infty}\int_{-\infty}^{\infty} d^4\left(a_{IA}/\varepsilon\right)|a_{IA}/\varepsilon,\, n_3\rangle\langle a_{IA}/\varepsilon,\, n_3|. \tag{4.57}$$

We also assume orthogonality relations

$$\langle a_{IA}, n_3 | a_{IA}', n_3'\rangle = \delta_{n_3 n_3'}\delta^4\left(a_{IA} - a_{IA}'\right);$$

$$\langle a_{IA}/\varepsilon, n_3 | a_{IA}'/\varepsilon, n_3'\rangle = \delta_{n_3 n_3'}\delta^4\left(\frac{a_{IA}}{\varepsilon} - \frac{a_{IA}'}{\varepsilon}\right). \tag{4.58}$$

In order to make the transformation from the position representation to the momentum representation, we need an expression for

$$\left\langle a_{IA}/\varepsilon,\ n_3 \middle| a_{IA}',n_3 \right\rangle \ . \tag{4.59}$$

To that end, let us consider the usual derivation of $\langle y_{IA} |\ p_{IA}\rangle$ for finite s. We have the commutation relation

$$p_{IA\mu}y_{IA\nu} - y_{IA\nu}p_{IA\mu} = ig_{\mu\nu}\,. \tag{4.60}$$

Substitution of the solutions (4.29) for $y_{IA}$ and $p_{IA}$ yields

$$\begin{aligned} &p_{IA\mu}y_{IA\nu} - y_{IA\nu}p_{IA\mu} \\ &= \frac{1}{2}\left[a_{IA\mu}\frac{a_{IA\nu}}{\varepsilon} - \frac{a_{IA\nu}}{\varepsilon}a_{IA\mu}\right]\exp\left[2\sqrt{\varepsilon}\ \omega_0 s\right] \\ &+ \frac{1}{2}\left[a_{IA\mu}\frac{b_{IA\nu}}{\varepsilon} - \frac{b_{IA\nu}}{\varepsilon}a_{IA\mu}\right] - \frac{1}{2}\left[b_{IA\mu}\frac{a_{IA\nu}}{\varepsilon} - \frac{a_{IA\nu}}{\varepsilon}b_{IA\mu}\right] \\ &- \frac{1}{2}\left[b_{IA\mu}\frac{b_{IA\nu}}{\varepsilon} - \frac{b_{IA\nu}}{\varepsilon}b_{IA\mu}\right]\exp\left[-2\sqrt{\varepsilon}\ \omega_0 s\right] = ig_{\mu\nu}\,. \end{aligned} \tag{4.61}$$

This implies the commutation relation $[a_{IA\mu}\ ,\ b_{IA\mu}] = i\varepsilon g_{\mu\nu}$, given in (4.30). As expected, in the asymptotic regions $s \to \pm\infty$, the leading terms of the operators $y_{IA}$ and $p_{IA}$ commute. However, it is in this region that $y_{IA}$ becomes singular and thus unmeasurable.

In the position representation, the operator $p_{IA}$ is represented by

$$\langle y_{IA}|p_{IA\mu} = i\frac{\partial}{\partial y_{IA}{}^{\mu}}\langle y_{IA}|. \tag{4.62}$$

Therefore, it follows that

$$p_{IA\mu}\langle y_{IA}|\ p_{IA}\rangle = \langle y_{IA}|p_{IA\mu}|p_{IA}\rangle = i\frac{\partial}{\partial y_{IA}{}^{\mu}}\langle y_{IA}|\ p_{IA}\rangle. \tag{4.63}$$

The solution to this differential equation has the form

$$\langle y_{IA}|\ p_{IA}\rangle = \text{const.} \times \exp\left[-iy_{IA}p_{IA}\right]. \tag{4.64}$$

We shall assume this expression holds for the present model both for finite s and for $s \to \pm\infty$, with the caveat that the constant of proportionality might be zero.

In analogy to (4.64), we expect the transformation function $\left\langle a_{IA}/\varepsilon,\ n_3 \middle| a_{IA}',n_3 \right\rangle$ to take the form

$$\left\langle a_{IA}/\varepsilon,\ n_3 \,\middle|\, a_{IA}',n_3 \right\rangle = \text{const.}\times\exp\left(-i\frac{a_{IA}}{\varepsilon}\cdot a_{IA}'\right). \tag{4.65}$$

However, consider the equation below (we use boldface to indicate an operator):

$$a_{IA\mu}'\left\langle a_{IA\mu}/\varepsilon,\ n_3 \,\middle|\, a_{IA\mu}',n_3 \right\rangle = \left\langle a_{IA\mu}/\varepsilon,\ n_3 \middle| \mathbf{a_{IA\mu}} \middle| a_{IA\mu}',n_3 \right\rangle. \tag{4.66}$$

Now, if $x=y$, we cannot infer $x/\varepsilon = y/\varepsilon$, but only[20]

$$\frac{x}{\varepsilon} = \frac{y}{\varepsilon} + c\delta(\varepsilon), \tag{4.67}$$

where c is unknown. Thus, dividing both sides of (4.66) by $\varepsilon$ yields

$$\left(\frac{a_{IA\mu}'}{\varepsilon} - \frac{a_{IA\mu}}{\varepsilon}\right)\left\langle a_{IA\mu}/\varepsilon,\ n_3 \,\middle|\, a_{IA\mu}',n_3 \right\rangle = c\,\delta(\varepsilon). \tag{4.68}$$

Multiplying the above by $\varepsilon$ then gives us

$$\left(a_{IA\mu}' - a_{IA\mu}\right)\left\langle a_{IA\mu}/\varepsilon,\ n_3 \,\middle|\, a_{IA\mu}',n_3 \right\rangle = 0. \tag{4.69}$$

The difficulty with this equation is that knowing a value for $a_{IA\mu}/\varepsilon$ does not tell us the value of $a_{IA\mu}$, and vice versa, so that, in general, the above equation imparts no information. There is one situation, however, that allows a conclusion to be drawn from (4.69). Although we can't know both $a_{IA\mu}/\varepsilon$ and $a_{IA\mu}$, we do know they have the same sign. Thus, if $a_{IA\mu}'$ and $a_{IA\mu}$ have opposite signs, it follows that

$$\left\langle a_{IA\mu}/\varepsilon,\ n_3 \,\middle|\, a_{IA\mu}',n_3 \right\rangle = 0, \qquad \left(a_{IA\mu}' \text{ and } a_{IA\mu} \text{ opposite sign}\right). \tag{4.70}$$

Similar conclusions can be reached for states and operators corresponding to $s \to -\infty$. Thus, for $s \to \pm\infty$, we shall express the transforms between position and momentum representations as follows:

$$\left|p_{IA}(\pm\infty),n_3\right\rangle = \frac{1}{(2\pi)^4}\left[\Theta\left(p_{IA0}(\pm\infty)\right)\int_0^\infty dy_{IA0}(\pm\infty) + \Theta\left(-p_{IA0}(\pm\infty)\right)\int_{-\infty}^0 dy_{IA0}(\pm\infty)\right]$$

$$\times\prod_{i=1}^{3}\left[\Theta\left(p_{IAi}(\pm\infty)\right)\int_0^\infty dy_{IAi}(\pm\infty) + \Theta\left(-p_{IAi}(\pm\infty)\right)\int_{-\infty}^0 dy_{IAi}(\pm\infty)\right]$$

$$\times e^{iy_{IA}p_{IA}}\left|y_{IA}(\pm\infty),n_3\right\rangle; \tag{4.71}$$

$$|y_{IA}(\text{"}\,4), n_3\rangle = \frac{1}{\acute{a}2\pi\acute{e}^{4}}\left[\Theta\,\acute{a}y_{IA0}(\text{"}\,4)\acute{e}\mathop{\mathrm{I}}\limits_{0}^{4} dp_{IA0}(\text{"}\,4) + \Theta\,\acute{a}-y_{IA0}(\text{"}\,4)\acute{e}\mathop{\mathrm{I}}\limits_{-4}^{0} dp_{IA0}(\text{"}\,4)\right]$$

$$\mathrm{H}\mathop{\mathrm{J}}\limits_{i=1}^{3}\left[\Theta\,\acute{a}y_{IAi}(\text{"}\,4)\acute{e}\mathop{\mathrm{I}}\limits_{0}^{4} dp_{IAi}(\text{"}\,4) + \Theta\,\acute{a}-y_{IAi}(\text{"}\,4)\acute{e}\mathop{\mathrm{I}}\limits_{-4}^{0} dp_{IAi}(\text{"}\,4)\right]$$

$$\mathrm{H}\,e^{-iy_{IA}p_{IA}}\,|\,p_{IA}(\text{"}\,4), n_3\rangle\,. \qquad (4.72)$$

It is important to note that these transformations hold because of the special coordinate frame induced by the harmonic-oscillator solutions and the n.b.c. In this frame, the quark vectors $y_{IA\mu}$ and $p_{IA\mu}$ are aligned.

## V. COMPOSITE-PARTICLE SCATTERING BY QUARK EXCHANGE

The nature of the harmonic-oscillator interaction allows the straightforward consideration of independent subgroups of quarks, or cluster decomposition, in the model of Sec. IV. We shall consider subgroups which have been "kinematically" decoupled from the remainder of the system, by setting for all I and A outside of the subgroup being considered:

$$\mathop{3}\limits_{I} a_{IA} = \mathop{3}\limits_{I} b_{IA} = 0, \quad \text{and} \quad \mathop{3}\limits_{I}\mathop{3}\limits_{A} a_{IA} b_{IA} = 0 \qquad (5.1)$$

The first relation comes from the transformation of coordinates (4.22). The second relation leaves the Hamiltonian (4.35) independent of those coordinates.

Two kinds of scattering problems will be examined, namely lepton-lepton and hadron-hadron scattering in four-quark clusters. The results are essentially the same as obtained in the first paper of Ref. 26, so the reader already familiar with the model may want to go directly to Sec. VI.

### A. Lepton-lepton scattering

The simplest description of particle scattering corresponds to a system of four lepton quarks, which we shall take to be the following:

$$Q_{IA} = \frac{1}{\varepsilon\sqrt{2M}}\left[a_{IA}\exp\ddot{a}\sqrt{\varepsilon}\,\omega_0 s\,\ddot{e} + b_{IA}\exp\ddot{a}-\sqrt{\varepsilon}\,\omega_0 s\,\ddot{e}\right] + \frac{1}{\sqrt{4NM}}\left[A_1 s + B_1\right]$$

$$+\left(-1\right)^{\frac{A-1}{2}}\frac{\cosh\left(\sqrt{\varepsilon}\,\omega_0 s\right)}{\sqrt{4NM}}\left[a_3^{\dagger}\exp\left(i\omega_0 s/\sqrt{\varepsilon}\right)+a_3\exp\left(-i\omega_0 s/\sqrt{\varepsilon}\right)\right], \quad (5.2)$$

where $I=1,\ 2;\ A=1,3$. From the condition $\sum_{I=1}^{N} y_{IA}=0$ and (5.1), it follows that,

$$a_{11}=-a_{21},\ \ a_{13}=-a_{23},\ b_{11}=-b_{21},\ \ b_{13}=-b_{23}. \quad (5.3)$$

In order to picture the scattering, it is helpful to consider a particular solution. Consider the case where

$$a_{110}<0,\ \ b_{110}<0,\ \ a_{230}<0,\ \ b_{130}<0. \quad (5.4)$$

The first two of these relations imply that the quark $Q_{11}$ turns around in the observer's time, so that at $t=-\infty$, the same quark appears in two different locations in space.[32] The other quarks at $t=-\infty$ are $Q_{23}$ and $Q_{13}$.

Consider now what quarks exist at $t=+\infty$. It follows from (5.3) and (5.4) that

$$a_{120}>0,\ \ \ b_{120}>0,\ \ \ a_{130}>0,\ \ b_{230}>0. \quad (5.5)$$

Thus, the quarks $Q_{23}$, $Q_{13}$, and $Q_{21}$ exist in the final state $t=+\infty$, with $Q_{21}$ occurring at two spatial locations also.

In order to describe two leptons in the initial state, we apply quark confinement relations as initial conditions at $t=-\infty$:

$$a_{11}=a_{23},\qquad b_{11}=b_{13}. \quad (5.6)$$

For the observer's final state at $t=+\infty$, using (5.3) and (5.6), we find

$$a_{21}=a_{13}\ \text{with}\ \ a_{210}=a_{130}>0,\ \ \text{and}\ b_{21}=b_{23}\ \ \text{with}\ \ \ b_{210}=b_{230}>0. \quad (5.7)$$

*In other words, the initial state implies that there are four quarks in the final state, also paired into two leptons.*

From (5.3), (5.6), and (5.7), it follows that

$$\sum_{I=1}^{2}\sum_{A=1,3} a_{IA}b_{IA}=0. \quad (5.8)$$

Hence, the Hamiltonian (4.35) becomes

$$H=\left(\omega/2\right)\left[-2a_2b_2+(1+i\varepsilon)^2\sum_{A=3}^{4}\left(a_A{}^{\dagger}a+a_Aa_A{}^{\dagger}\right)\right]. \quad (5.9)$$

Consider now the leptons formed in the asymptotic regions $s\to\pm\infty$. With the aid of the quark solutions (5.2), we obtain

$$Q_1^{+L} \equiv \tfrac{1}{2}(Q_{11}+Q_{23})\big|_{s\to+\infty} \sim (1/\varepsilon)\,(2M)^{-1/2}\, a_{11}\exp(\sqrt{\varepsilon}\,\omega_0 s),$$

$$Q_2^{+L} \equiv \tfrac{1}{2}(Q_{21}+Q_{13})\big|_{s\to+\infty} \sim -(1/\varepsilon)\,(2M)^{-1/2} a_{11}\exp(\sqrt{\varepsilon}\,\omega_0 s),$$

$$Q_1^{-L} \equiv \tfrac{1}{2}(Q_{11}+Q_{13})\big|_{s\to-\infty} \sim (1/\varepsilon)\,(2M)^{-1/2} b_{11}\exp(-\sqrt{\varepsilon}\,\omega_0 s),$$

$$Q_2^{-L} \equiv \tfrac{1}{2}(Q_{21}+Q_{23})\big|_{s\to-\infty} \sim -(1/\varepsilon)\,(2M)^{-1/2} b_{11}\exp(-\sqrt{\varepsilon}\,\omega_0 s). \qquad (5.10)$$

The leptons all have internal states described by the oscillatory solution for $W_3$:

$$\begin{aligned} q_l^{\pm L}(s) &\equiv (2N)^{-1/2} W_3 \\ &= (4NM)^{-1/2}\cosh\left(\sqrt{\varepsilon}\,\omega_0 s\right)\left[a_3^{\dagger}\exp(i\omega_0 s/\varepsilon) + a_3\exp(-i\omega_0 s/\varepsilon)\right]. \end{aligned} \qquad (5.11)$$

The lepton momenta are obtained from (4.25) and (4.29):

$$P_{1,2}^{+L} \equiv (P_{11}+P_{23})\big|_{s\to+\infty} \sim (2M)^{1/2} a_{11}\exp(\sqrt{\varepsilon}\,\omega_0 s),$$

$$P_{1,2}^{+L} \equiv (P_{21}+P_{13})\big|_{s\to+\infty} \sim -(2M)^{1/2} a_{11}\exp(\sqrt{\varepsilon}\,\omega_0 s),$$

$$P_{1,2}^{-L} \equiv (P_{11}+P_{13})\big|_{s\to-\infty} \sim -(2M)^{1/2} b_{11}\exp(-\sqrt{\varepsilon}\,\omega_0 s),$$

$$P_{1,2}^{-L} \equiv (P_{21}+P_{23})\big|_{s\to-\infty} \sim (2M)^{1/2} b_{11}\exp(-\sqrt{\varepsilon}\,\omega_0 s). \qquad (5.12)$$

From (5.12) and the natural boundary conditions (4.42), we see that the leptons all have the same mass, or, letting $\varepsilon \to 0$,

$$m_{n_3}^{\;2} = (M/2N)(n_3+2). \qquad (5.13)$$

It is apparent from (5.12) that four-momentum is conserved:

$$P_1^{+L} + P_2^{+L} + P_1^{-L} + P_2^{-L} = 0. \qquad (5.14)$$

Since the orbital angular momentum of the leptons is zero, and the internal states are the same, it follows also that angular momentum is conserved.

Thus, the following picture of lepton-lepton scattering emerges. In the asymptotic regions of s, leptons are formed which behave as free particles. Scattering occurs by quark exchange and in the forward or backward directions only. Since the leptons are identical, one can’t discern that any scattering has taken place. In other words, the leptons behave as if they are point particles and do not interact in this “strong interaction” harmonic-oscillator model.

## B. Hadron-hadron scattering

Let us now consider a decoupled subgroup of four hadron quarks described by the position four-vectors

$$Q_{IA} = \frac{1}{\varepsilon\sqrt{2M}} \left\{ \left[ a_{IA} + (2N)^{-1/2} a_2 \right] \exp(\sqrt{\varepsilon}\,\omega_0 s) + \left[ b_{IA} + (2N)^{-1/2} b_2 \right] \exp(-\sqrt{\varepsilon}\,\omega_0 s) \right.$$

$$\left. + (-1)^{\frac{A-2}{2}} \frac{\cosh\sqrt{\varepsilon}\,\omega_0 s}{\sqrt{4MN}} \left[ a_4^{\dagger} \exp\left( i\omega_0 s/\sqrt{\varepsilon} \right) + a_4 \exp\left( -i\omega_0 s/\sqrt{\varepsilon} \right) \right] \right\}, \qquad (5.15)$$

where $I = 1, 2$; $A = 2, 4$. The decoupling from the remainder of the system implies

$$\sum_{I=1}^{2} a_{IA} = \sum_{I=1}^{2} b_{IA} = 0, \qquad A = 2, 4. \qquad (5.16)$$

Following the similar line of investigation carried out for the lepton scattering, we shall suppose that at $t = -\infty$, quark confinement implies

$$a_{12} = a_{24}, \text{ and } a_{120} < 0; \quad b_{12} = b_{14} \text{ and } b_{120} < 0. \qquad (5.17)$$

In addition, the quark-antiquark assumption for the initial state implies $a_2 = b_2$. It follows that the following hadrons exist at $s = \pm\infty$:

$$Q_1^{+H} \equiv \tfrac{1}{2}(Q_{12} + Q_{24})\big|_{s \to +\infty} \approx (1/\varepsilon)(2M)^{-1/2}\left[ a_{12} + (2N)^{-1/2} a_2 \right] \exp(\sqrt{\varepsilon}\,\omega_0 s),$$

$$Q_2^{+H} \equiv \tfrac{1}{2}(Q_{22} + Q_{14})\big|_{s \to +\infty} \approx (1/\varepsilon)(2M)^{-1/2}\left[ -a_{12} + (2N)^{-1/2} a_2 \right] \exp(\sqrt{\varepsilon}\,\omega_0 s),$$

$$Q_1^{-H} \equiv \tfrac{1}{2}(Q_{12} + Q_{14})\big|_{s \to -\infty} \approx (1/\varepsilon)(2M)^{-1/2}\left[ b_{12} + (2N)^{-1/2} a_2 \right] \exp(-\sqrt{\varepsilon}\,\omega_0 s),$$

$$Q_2^{-H} \equiv \tfrac{1}{2}(Q_{22} + Q_{24})\big|_{s \to -\infty} \approx (1/\varepsilon)(2M)^{-1/2}\left[ -b_{12} + (2N)^{-1/2} a_2 \right] \exp(-\sqrt{\varepsilon}\,\omega_0 s). \qquad (5.18)$$

The hadron internal states are described by

$$q_I^{\pm H}(s) = \frac{\cosh\sqrt{\varepsilon}\,\omega_0 s}{\sqrt{4NM}} \left[ a_4^{\dagger} \exp\left( i\omega_0 s/\sqrt{\varepsilon} \right) + a_4 \exp\left( -i\omega_0 s/\sqrt{\varepsilon} \right) \right]. \qquad (5.19)$$

The hadron momenta are

$$P_1^{+H} \equiv (P_{12} + P_{24})\big|_{s \to +\infty} \approx (2M)^{1/2}\left[ a_{12} + (2N)^{-1/2} a_2 \right] \exp(\sqrt{\varepsilon}\,\omega_0 s),$$

$$P_2^{+H} \equiv (P_{22} + P_{14})\big|_{s \to +\infty} \approx (2M)^{1/2}\left[ -a_{12} + (2N)^{-1/2} a_2 \right] \exp(\sqrt{\varepsilon}\,\omega_0 s),$$

$$P_1^{-H} \equiv (P_{12}+P_{14})\big|_{s\to\infty} \to -(2M)^{1/2}\left[b_{12}+(2N)^{-1/2}a_2\right]\exp(-\sqrt{\varepsilon}\,\omega_0 s),$$

$$P_2^{+H} \equiv (P_{22}+P_{24})\big|_{s\to\infty} \to -(2M)^{1/2}\left[-b_{12}+(2N)^{-1/2}a_2\right]\exp(-\sqrt{\varepsilon}\,\omega_0 s). \qquad (5.20)$$

From the n.b.c. and (5.20), we obtain the mass-shell constraints

$$\left(P_I^{\pm H}\right)^2 = m_{n_4}{}^2 = (M/2N)(n_4+2). \qquad (5.21)$$

Examination of (5.20) shows that total four-momentum is conserved:

$$P_1^{+H}+P_2^{+H}+P_1^{-H}+P_2^{-H}=0. \qquad (5.22)$$

The initial conditions (5.17) dictate the configuration of the scattering. In the case just considered, two quarks, $Q_{12}$ and $Q_{22,}$ turn around in time. A different configuration can be chosen (by different initial conditions), in which quarks $Q_{14}$ and $Q_{24}$ turn around in time. The two configurations are similar in nature, but differ in the assignment of quarks in the initial state. In addition to these configurations, a third type is possible where none of the four quarks turn around in time (this third configuration is not an option in lepton-lepton scattering). All three configurations can be described by the solutions (5.18) and (5.20) if the appropriate signs are chosen for $\left[\pm a_{110}+(2N)^{-1/2})a_{20}\right]$ and $\left[\pm b_{110}-(2N)^{-1/2})a_{20}\right]$.

Thus, we arrive at the following picture of hadron-hadron scattering. It is similar to lepton-lepton scattering in the following ways: (1) Scattering takes place by quark exchange; (2) quark-antiquark confinement in the initial state (observer's time $t=-\infty$) implies quark-antiquark confinement in the final state ($t=+\infty$); (3) total four-momentum is conserved (a result of the antiquark assumptions of Sec.IV); (4) the hadrons all have zero orbital angular momentum, and equal internal angular momentum. However, there is a crucial difference between lepton and hadron scattering. Hadron scattering can occur at angles other than $0^o$ and $180^o$. Shortly we will calculate the scattering amplitudes, when it will be clear that the constant $a_2$ plays the role of an impact parameter.

In summary, the assumption of quark confinement in the initial state for both lepton and hadron scattering implies confinement in the final state. The leptons, however, can scatter in the forward/backward directions only, tantamount to no scattering at all. The hadrons, on the other hand, can scatter in other directions as well. We shall now formulate scattering amplitudes for hadron-hadron scattering.

## C. State vectors for hadron scattering

A complete set of commuting operators for the four quarks includes (to first order)

$$a_{IA\mu},\ b_{IA\mu},\ a_{2\mu},\ b_{2\mu},\ n_3,\ \text{and}\ n_4,\ I=1,\ 2\ \text{and}\ A=2,4, \tag{5.23}$$

where $n_A \equiv -a_A{}^{\dagger} a_A$. The eigenvalues of these operators can be used to label the eigenstates of the Hamiltonian (4.40). We shall require that the initial states must also satisfy the following conditions:

(1) From the transformation $Q_{IA} \equiv y_{IA} + N^{-1/2} \langle \delta W \rangle_A$ and cluster decomposition,

$$\sum_{I=1}^{2} a_{IA}\left|\Psi\right\rangle = \sum_{I=1}^{2} b_{IA}\left|\Psi\right\rangle = 0. \tag{5.24}$$

(2) The quark-antiquark assumption

$$a_2 = b_2. \tag{5.25}$$

(3) From the n.b.c. (4.43),

$$\left[a_{IA} + [2N]^{-1/2} a_2\right]^2 \left|\Psi\right\rangle = \left[b_{IA} + [2N]^{-1/2} b_2\right]^2 \left|\Psi\right\rangle = \frac{1}{4N}\,[n_4 + 2]\left|\Psi\right\rangle, \quad \text{N=2, 4;} \tag{5.26}$$

$$\begin{aligned} &a_4{}^2\left|\Psi\right\rangle; \quad a_{IA}\cdot a_4\left|\Psi\right\rangle = b_{IA}\cdot a_4\left|\Psi\right\rangle = 0, \quad A=2,\ 4;\\ &a_2\cdot a_4\left|\Psi\right\rangle = 0; \quad b_2\cdot a_4\left|\Psi\right\rangle = 0; \end{aligned} \tag{5.27}$$

(4) From the constraint $H \approx 0$ and the Hamiltonian (4.40), it follows that, to first order,

$$\left[\sum_{I=1}^{2}\ \sum_{A=2,4} a_{IA} b_{IA} + a_2 b_2\ + \sum_{A=3}^{4} (n_A + 2)\right]\left|\Psi\right\rangle = 0. \tag{5.28}$$

We shall denote the eigenvectors which satisfy conditions (5.27) as

$$\left|\Psi\right\rangle \equiv \left|a_{IA}, b_{IA}, a_2, b_{2,} n_3, n_4\right\rangle, \quad I=1,\ 2,\ \text{and}\ A=2,4. \tag{5.29}$$

The conditions (5.26) and (5.28) will be imposed separately as delta functions.

Note, for the eigenvector above, $n_3$ and $n_4$ can take on negative values. Negative mass-squared composites are not ruled out. We ask only that they not be coupled to positive mass-squared particles.

## D. Interpretation of the physical state vectors

The commuting operators (5.23) result in a complete set of simultaneous eigenstates, but the conditions (5.24)-(5.28) reduces the allowed initial states to a subgroup of those states. In the usual formulation of scattering problems, one generally employs two complete sets of free-particle state vectors to represent the "incoming" and "outgoing" states of the system. One goes from the "in" basis to the "out" basis by means of a unitary transformation, or S-matrix. A particular transition is characterized by an S-matrix element symbolized, for example, by

$$\left\langle \xi_{out} \middle| \xi_{in}{'} \right\rangle = \left\langle \xi_{out} \middle| S \middle| \xi_{out}{'} \right\rangle . \tag{5.30}$$

In the present case, however, *the S-matrix formulation does not apply.* In the harmonic-oscillator strong-interaction model, the state vector contains the eigenvalues of all the operators corresponding $t = \pm\infty$ in the observer's time. Therefore, it is meaningless to speak of the transition of the state describing the hadrons at $t = -\infty$ to a state at $t = +\infty$. A particular solution must be characterized by eigenvalues specifying what hadrons exist and what their behavior is at both $t = \pm\infty$.

At first, this seems to contradict experiment. For example, repeating a scattering experiment with identical physical initial conditions yields a distribution of scattering angles. Taking a closer look at the model reveals there is no contradiction, however. This is because physical measurements at $t = -\infty$ do not determine all of the eigenvalues of the operators in (5.23). Further, the addition of the physical measurements of the final state still does not determine all the eigenvalues. As we shall now see, the scattering amplitude reflects this indeterminacy by including a sum over the unknown parameters.

### E. Hadron-hadron scattering amplitudes

The physical initial conditions in the observer's time $t = -\infty$ consist of the incoming hadron momenta and masses. These do not determine which of the four quarks are involved. Thus the state vector must be expressed as a linear combination of state corresponding to all possible configurations that might arise. We shall assume that these states have equal probability of occurring.

Denote the incoming hadron momenta as $P_1$ and $P_2$. The three possible quark configurations will be defined as the s-channel, t-channel, and u-channel configurations. For the s-channel, put

$$P_1 \equiv P_1^{-H} = -(2M)^{1/2}\left[b_{12} + (2N)^{-1/2} a_2\right];$$
$$P_2 \equiv P_2^{-H} = -(2M)^{1/2}\left[-b_{12} + (2N)^{-1/2} a_2\right]. \tag{5.31}$$

These momenta are taken from (5.20) in the limit $s \to -\infty$ and $\varepsilon \to 0$. The incoming state vector at $t = -\infty$ becomes

$$\left|\Psi^{s-ch}_{IN}\right\rangle = \sum_{n_3=-\infty}^{\infty} \prod_{I=1}^{2} \prod_{A=2,4} \int d^4 a_{IA} \int d^4 b_{IA} \int d^4 a_2 \ \delta\left(a_2{}^2 + n_3 + n_4 + 4\right)$$

$$\times\ \delta\left(P_1 + (2M)^{1/2}\left[b_{12} + (2N)^{-1/2} a_2\right]\right)\ \delta\left(P_2 + (2M)^{1/2}\left[-b_{12} + (2N)^{-1/2} a_2\right]\right)$$

$$\times \left|a_{IA}, b_{IA}, a_2, b_2 = -a_2, n_3, n_4\right\rangle . \qquad (5.32)$$

Recall that the state $\left|a_{IA}, b_{IA}, a_2, b_2 = -a_2, n_3, n_4\right\rangle$ obeys conditions (5.25) and (5.27). The conditions (5.26) and (5.28), appear in the delta function above.

The conditions (5.24) - (5.26) imply that the final state consists also of two composites, each of mass $M_4$. Label the composite momenta at $t = +\infty$ as $P_3$ and $P_4$. Then, from (5.20), it follows that we can put

$$P_3 \equiv P_1^{+} = (2M)^{1/2}\left[a_{12} + (2N)^{-1/2} a_2\right];$$

$$P_4 \equiv P_2^{+} = (2M)^{1/2}\left[-a_{12} + (2N)^{-1/2} a_2\right]. \qquad (5.33)$$

The outgoing state in the s-channel is thus

$$\left\langle\Psi^{s-ch}_{OUT}\right| = \sum_{n_3=-\infty}^{\infty} \prod_{I=1}^{2} \prod_{A=2,4} \int d^4 a_{IA} \int d^4 b_{IA} \int d^4 a_2 \int d^4 b_2 \ \delta\left(a_2{}^2 - n_3 - n_4 - 4\right)$$

$$\times\ \delta\left(P_3 - (2M)^{1/2}\left[a_{12} + (2N)^{-1/2} a_2\right]\right)\ \delta\left(P_4 - (2M)^{1/2}\left[-a_{12} + (2N)^{-1/2} a_2\right]\right)$$

$$\times\ \left\langle a_{IA}, b_{IA}, a_2, b_2, n_3, n_4\right| . \qquad (5.34)$$

Calculating the overlap of these two state vectors yields

$$\left\langle\Psi^{s-ch}_{OUT}\middle|\ \Psi^{s-ch}_{IN}\right\rangle = \delta\left(P_1 + P_2 + P_3 + P_4\right) \prod_{I=1}^{4} \delta\left(P_I{}^2 - M_4{}^2\right)$$

$$\times \sum_{n=-\infty}^{\infty} \delta\left(N(4M)^{-1}\left(P_1 + P_2\right)^2 - n\right). \qquad (5.35)$$

The final state is characterized by $b_2=a_2$, and we may identify the composites as hadrons.

The t-channel and u-channel contributions are treated similarly,[32] leading to a hadron-hadron scattering amplitude given by

$$A(s,t,u) = N\ \left[\left\langle\Psi^{s-ch}_{OUT}\middle|\ \Psi^{s-ch}_{IN}\right\rangle + \left\langle\Psi^{t-ch}_{OUT}\middle|\ \Psi^{t-ch}_{IN}\right\rangle + \left\langle\Psi^{u-ch}_{OUT}\middle|\ \Psi^{u-ch}_{IN}\right\rangle\right]$$

$$= N \prod_{I=1}^{4} \delta\left(P_I{}^2 - M_4{}^2\right)\ \delta\left(P_1 + P_2 + P_3 + P_4\right)\left[D(s) + D(t) + D(u)\right], \qquad (5.36)$$

where $N$ is a proportionality constant and

$$D(z) \equiv \sum_{n=-\infty}^{\infty} \delta\left(\left(N/4M\right)\left(z-n\right)\right), \tag{5.37}$$

and

$$s=\left(P_1+P_2\right)^2; \quad t=\left(P_1+P_3\right)^2; \quad u=\left(P_1+P_4\right)^2. \tag{5.38}$$

The variable s here should not be confused with the evolution parameter.

In the center-of-energy (CE) system,

$$s=\left(E_{CE}\right)^2; \quad t=-2q^2\left(1-\cos\theta\right); \quad u=-q^2\left(1+\cos\theta\right), \tag{5.39}$$

where $E_{CE}$ is the energy, q is the three-momentum, and $\theta$ is the scattering angle, all taken in the CE system.

The delta function in (5.37) can be recast in the interesting form

$$\delta(z-n)=\lim_{\varepsilon\to 0} -\frac{1}{2\pi i}\left[\frac{1}{z-n+i\varepsilon}-\frac{1}{z-n-i\varepsilon}\right]. \tag{5.40}$$

Thus, for finite $\varepsilon$, the scattering amplitude has simple poles at $z=n \pm i\varepsilon$ with constant residues. This can be compared to amplitudes for various dual models, such as the Veneziano model,[24] where the simple poles in s, for example, have residues which are polynomials in t. In the present simple version of the harmonic oscillator, the orbital angular momenta of the hadrons are zero. Thus we expect a constant residue.

## VI. QUARK ELECTROWEAK MODEL

### A. String-like Electroweak Lagrangian

In Sec. I, a general string-like Lagrangian was introduced:

$$L=-\left\{-\left[\mathbf{F}(\mathbf{Q})\right]^2 \dot{\mathbf{Q}}^2+\left[\dot{\mathbf{Q}}\cdot\mathbf{F}(\mathbf{Q})\right]^2\right\}^{1/2}, \tag{6.1}$$

where **F**(**Q**) depends on the quark coordinates **Q**(s). We have assumed that at certain energies and in certain domains of space-time, we can approximate **F**(**Q**) by particular functions. For example, in a strong-interaction domain, we put **F**(**Q**)=M**HQ**, with the harmonic-oscillator matrix **H** given by (4.8). In this section, we shall set **F**(**Q**)=**A**(**Q**) and consider an expression for **A**(**Q**) suitable for the description of electroweak interactions. In choosing **A**(**Q**), one's first impulse is to base the choice on the classical Wheeler-Feynman model of action-at-a-distance electrodynamics.[16,17]

However, this does not yield Feynman amplitudes in general.

Interestingly, as we shall show, the Wheeler-Feynman potential can be regarded as a four-dimensional generalization of a simple one-dimensional problem, namely, the nonrelativistic collision of two point particles. This one-dimensional problem can be formulated in other ways. We choose one of these alternatives as the basis of the relativistic generalization of the potential. First-order perturbation theory is then applied to an example of spinless lepton-lepton elastic scattering. The unperturbed state is the asymptotic harmonic-oscillator state for two leptons from different quark clusters. *Thus, the electroweak interaction between leptons (and between hadrons) arises because of interactions between the constituent quarks of one lepton with the constituent quarks of a second lepton.*

The basis of the Wheeler-Feynman action-at-a-distance model is the Lorentz-invariant action devised by Fokker:[17]

$$I_F \equiv -\sum_i \dot{q}_i^{\,2}(s)ds - \sum_i e_i \int_{-\infty}^{\infty} A_i[q(s)]\,\dot{q}_i(s)ds, \tag{6.2}$$

where

$$A_{i\mu}(q_i) \equiv \sum_{i<j} e_j \int_{-\infty}^{\infty} \delta\left([q_i(s)-q_j(s')]^2\right)\dot{q}_{j\mu}(s')ds'. \tag{6.3}$$

We replace the action above by the following:

$$I = \int_{-\infty}^{\infty} L(s)ds, \tag{6.4}$$

where

$$L = -\left\{-[\mathbf{A}(\mathbf{Q})]^2\dot{\mathbf{Q}}^2 + [\mathbf{A}(\mathbf{Q})\cdot\dot{\mathbf{Q}}]^2\right\}^{1/2}, \tag{6.5}$$

and $\mathbf{A}(\mathbf{Q})$ remains to be specified.

The Lagrangian (6.5) yields the primary constraints

$$\Phi_1 \equiv \mathbf{P}^2 + \mathbf{A}^2 \approx 0, \quad \text{and} \quad \Phi_2 \equiv \mathbf{A}\cdot\mathbf{P} \approx 0. \tag{6.6}$$

We shall *assume* consistency is maintained with a gauge choice (see the discussion of the n.b.c. and constraints in Sections III and IV)

$$v_1 = \frac{1}{2} v_2 = \frac{1}{2}. \tag{6.7}$$

The quantized Dirac Hamiltonian then becomes

$$H = \frac{1}{2}\left[\mathbf{P}^2 + \mathbf{A}\cdot\mathbf{P} + \mathbf{P}\cdot\mathbf{A} + \mathbf{A}^2\right]. \tag{6.8}$$

The Hamiltonian displays the familiar “minimal interaction” substitution **P**+**A** for **P** in a Hamiltonian for a free particle.

If, to first order, we neglect the term $\mathbf{A}^2$, we can write

$$H = H_O + H_I, \tag{6.9}$$

with

$$H_O \equiv \frac{1}{2}\mathbf{P}^2 = \frac{1}{2}\sum_{I=1}^{N}\sum_{A=1}^{4} P_{IA}{}^2, \tag{6.10}$$

and

$$\begin{aligned} H_I &= \frac{1}{2}\left[\mathbf{A}\cdot\mathbf{P} + \mathbf{P}\cdot\mathbf{A}\right] \\ &= \frac{1}{2}\sum_{I=1}^{N}\sum_{A=1}^{4}\left[A_{IA}(Q_{IA})P_{IA} + P_{IA}A_{IA}(Q_{IA})\right] \end{aligned} \tag{6.11}$$

## B. The perturbation method

In the Schrödinger Picture, assume the state vector obeys equations of motion

$$H\left|\Psi(s)\right\rangle = -i\frac{\partial}{\partial s}\left|\Psi(s)\right\rangle. \tag{6.12}$$

If H is constant in s, the equation can be integrated to give

$$\left|\Psi(s)\right\rangle = \text{const.} \times e^{iHs}\left|\Psi\right\rangle. \tag{6.13}$$

The solutions in the case of a parametrically invariant action must also satisfy

$$H\left|\Psi(s)\right\rangle \approx 0. \tag{6.14}$$

Thus, there is no evolution of the system as a whole in s. However, we shall again rely on the parameter ε to circumvent this feature and provide a visualization of the process as a function of s.

Assume that H can be expressed as $H=H_0+H_I$, where $H_I$ is the perturbation term which is very small and almost vanishes at $s = \pm\infty$, that is,

$$H_I\left|\Psi(\pm\infty)\right\rangle \approx 0. \tag{6.15}$$

Then, (6.14) implies we can take

$$H_O\left|\Psi(\pm\infty)\right\rangle \approx 0. \tag{6.16}$$

In the Interaction Picture, a state $\left|\Phi(s)\right\rangle$ is defined through the relation

$$\left|\Psi(s)\right\rangle \equiv e^{iH_O s}\left|\Phi(s)\right\rangle. \tag{6.17}$$

Providing $H_0$ is nearly constant in s, the state $\left|\Phi(s)\right\rangle$ obeys, to first order, the equation of motion

$$\mathcal{H}_I(s)\left|\Phi(s)\right\rangle = -i\frac{\partial}{\partial s}\left|\Phi(s)\right\rangle, \tag{6.18}$$

where $\mathcal{H}_I(s)$ is defined as

$$\mathcal{H}_I(s) \equiv e^{-iH_O s} H_I e^{iH_O s}. \tag{6.19}$$

The equation of motion (6.18) can be integrated to get

$$\left|\Phi(s)\right\rangle = \left|\Phi(-S)\right\rangle - i\int_{-S}^{s} \mathcal{H}_I(s')\left|\Phi(s')\right\rangle ds'. \tag{6.20}$$

By iterating the equation, higher order terms may be obtained in approximating $\left|\Phi(s)\right\rangle$. Setting s=S in (6.20), we obtain to first order

$$\left|\Phi(S)\right\rangle \approx \left|\Phi(-S)\right\rangle - i\int_{-S}^{S} \mathcal{H}_I(s)ds\left|\Phi(-S)\right\rangle. \tag{6.21}$$

### C. The unperturbed states: Asymptotic harmonic-oscillator solutions

We wish to match the initial state quarks of the unperturbed Hamiltonian with asymptotic quark solutions generated by two different four-quark clusters in the strong-interaction domain. The quarks are “confined” pairwise into composite particles. Recall that quark confinement resulted from assumed *initial conditions*, not because of any *force* between the quarks. The asymptotic solutions of Sec. IV describe quarks which are oscillating about a straight-line trajectory in space-time. The frequency of oscillation tends to infinity as $\varepsilon \to 0$. In that limit, the quarks become free “strings.”

Let us review the harmonic-oscillator solutions considered in Sec. IV and Sec.V. Clusters of four quarks, either all lepton or all hadron quarks, are “kinematically decoupled” from the remainder of the system (cluster decomposition). In each cluster, the four quarks pair up into two leptons (or two hadrons) in each of the asymptotic regions $s \to \pm\infty$. *Although s does not correspond to the observer’s time, two leptons (hadrons) also exist in of the observer’s times* $t = \pm\infty$.

The composite particles generated in this way carry no “memory” of the interactions that produced them, and furthermore two composites coming from different clusters are uncorrelated *except in regard to their internal states and in regard to the observer's time.* The internal states are correlated because the cluster decomposition occurs for a given state of the total harmonic-oscillator system. Similar to particles associated with a given quantum field, the quarks all "share" the corresponding oscillator operators $a_A$ and $a_A{}^{\dagger}$. As a result, for a given solution, all leptons, and all hadrons, have the same mass and internal angular momentum. Thus, for the perturbation calculations, the initial states we shall consider will be two identical leptons or identical two hadrons. However, continuing in the spirit of simplification, we shall treat the composites as identifiable, i.e., we shall neglect statistics.

The parameter $\varepsilon$ plays several critical roles in the model of this paper. We have already seen in Sec. IV how it leads to spontaneous symmetry breaking yielding leptons and hadrons with different properties. It was also demonstrated that $\varepsilon$ is responsible for eliminating any background frame of reference for the system when it is expressed in terms of the "fundamental" coordinates $x_{IA}$ (see Sec. IV-F). For the system of coordinates $Q_{IA}$, obtained from the $x_{IA}$ by the transformation (4.5), the lack of background reference frame exhibits itself in a somewhat different way. The composite position operators are proportional to $1/\varepsilon$, but in compensation, the "interaction" volume of space-time, defined by the quark position vectors for finite s, shrinks to zero. The composites effectively interact by colliding. However, after the limits are taken, i.e., s 6 " 4 and $\varepsilon$ 6 0, the composite momenta become finite and well-defined. We have implicitly assumed the observer is somehow "outside" of the reference frame, but assigns "initial" and "final" momenta to colliding composites. (Recall that for the quark clusters considered, the observer will always see two composites in both the initial and final states t = " 4.) In real life, of course, the composites leave "tracks" which are measured by some kind of device, such as a liquid in which bubbles form as the particle passes through.

Returning now to the question of a background reference frame. We have seen that the harmonic-oscillator model of Secs. IV and V cannot be used to infer the spatial location of the final state composites in the observer's frame of reference. In other words, we are free to assign initial spatial position and momentum three-vectors, in the wave packet sense, to two composites "generated" by two different clusters. Shortly, we will discuss how this affects the transforms between position and momentum states appearing in (4.71) and (4.72).

With these preparatory remarks, let us now consider the electroweak scattering of two free leptons. Each lepton in the initial state is composed of two quarks and but generated from different clusters. As we noted earlier, the leptons are characterized by the eigenvalues of their momenta and but "share" the same eigenvalue of $n_3$. We reset the "clock" to $s=-\infty$ for each particle. In the Heisenberg picture, the quark momentum vectors for the asymptotic harmonic-oscillator regions $s\to\pm\infty$ are obtained from (4.25), (4.29) and (4.37), reproduced below:

$$P_{IA}=p_{IA}+N^{-1/2}\langle\delta\mathbf{p}\rangle_A,\ \ A=1,\,3; \tag{4.25}$$

$$p_{IA}=(M/2)^{1/2}\left[a_{IA}\exp(\varepsilon\omega s)-b_{IA}\exp(-\varepsilon\omega s)\right]; \tag{4.29}$$

$$p_3=i(M/2)^{1/2}\cosh\left(\sqrt{\varepsilon}\,\omega_0 s\right)\left[a_3^{\dagger}\exp\left(i\omega_0 s/\sqrt{\varepsilon}\right)-\ a_3\exp\left(-i\omega_0 s/\sqrt{\varepsilon}\right)\right]. \tag{4.37}$$

(Recall that $p_1$ vanishes. Also, note that to define a physical momentum, it is sometimes necessary to replace $P_{IA}$ by $-P_{IA}$. See, for example, the conservation of momentum relation (5.14)).

For the perturbation calculations, we shall work in the Schrödinger Picture (S.P.). Let us first transform the s-dependent H.P. operators to the operators in the S.P. For a given cluster, this is accomplished by the transformation

$$O\equiv\exp(-iHs)\,O(s)\,\exp(iHs), \tag{6.22}$$

where H is the redefined Hamiltonian of (4.40).

Thus, in the respective clusters in which they were generated, the quark four-vectors and momenta become the constant operators

$$Q_{IA}=(2M)^{-1/2}\left[(1/\varepsilon)\ (a_{IA}+b_{IA})+(-1)^{\frac{A-1}{2}}\ \left(2\sqrt{2N}\right)^{-1}(a_3^{\dagger}+a_3)\right]; \tag{6.23}$$

$$P_{IA}=(M/2)^{1/2}\left[(\ a_{IA}-b_{IA})+i\,(-1)^{\frac{A-1}{2}}\ \left(2\sqrt{2N}\right)^{-1}(a_3^{\dagger}-a_3)\right]. \tag{6.24}$$

The natural boundary conditions (n.b.c.) imply

$$P_{IA}^{\ 2}\to 0,\quad \text{as } s\to\pm\infty. \tag{6.25}$$

As we noted in Sec. IV, these expressions imply that the composite momenta (composites exist only for $s\to\pm\infty$) are aligned with the position four-vectors. However, we wish to consider the leptons in an observer's frame, where we are permitted to choose the position arbitrarily. Therefore we shall rewrite the operators as

$$Q_{IA}=q_{IA}+(-1)^{\frac{A-1}{2}}\ (2M)^{-1/2}\left(2\sqrt{2N}\right)^{-1}(a_3^{\dagger}+a_3), \tag{6.26}$$

$$P_{IA}=p_{IA}+i\,(-1)^{\frac{A-1}{2}}\ (M/2)^{1/2}\left(2\sqrt{2N}\right)^{-1}(a_3^{\dagger}-a_3); \tag{6.27}$$

and

$$\left[q_{IA\mu},p_{IA\nu}\right]=-ig_{\mu\nu}. \tag{6.28}$$

In the above, we have assumed the description of the internal state of the lepton is unaffected by the choice of frame for the composite particle. We continue to assume that the coordinates $q_{IA}$ are singular; that is,

$$q_{IA} \sim 1/\varepsilon. \tag{6.29}$$

Thus, it is necessary to work in the momentum representation $|p_{IA}(s),\ n_3\rangle$. We can define a position representation $|q_{IA}(s),\ n_3\rangle$ only for $\varepsilon$ finite and when position representations are summed over as intermediate states in a matrix element.

The completeness relations and transforms in Sec. IV hold in the special frame where the position and momentum vectors are aligned. These must then be modified for the reference frame of the lepton-lepton electroweak interaction.

Since the natural boundary conditions are part of the dynamical system, we assume, as we did in Sec. IV, the completeness relations

$$1 = \sum_{n=-\infty}^{\infty} \int_{-\infty}^{\infty} d^4 q_{IA} \, |q_{IA},\ n\rangle \langle q_{IA},\ n|, \tag{6.30}$$

$$1 = \sum_{n=-\infty}^{\infty} \int_{-\infty}^{\infty} d^4 p_{IA} \, |p_{IA},\ n\rangle \, \delta\left(p_{IA}{}^2 - m_n{}^2\right) \langle p_{IA},\ n|, \tag{6.31}$$

where $m_n{}^2$ is given by (5.13). Analogous to the states defined in Sec. IV, the eigenstates $|p_{IA}, n_3\rangle$ are defined as satisfying the n.b.c.

$$a_3{}^2 \, |p_{IA}, n_3\rangle = 0, \qquad p_{IA} \cdot a_3 \, |p_{IA}, n_3\rangle = 0. \tag{6.32}$$

The delta function in (6.31) comes from the remaining n.b.c. in (4.42), which is an eigenvalue equation.

The momentum and position states are assumed to satisfy orthogonality relations

$$\langle p_{IA}, n_3 | \, p_{IA}{}', n_3{}' \rangle = \delta_{n_3', n_3} \, \delta^4\left(p_{IA} - p_{IA}{}'\right); \tag{6.33}$$

$$\langle q_{IA}, n_3 | \, q_{IA}{}', n_3{}' \rangle = \delta_{n_3', n_3} \, \delta^4\left(q_{IA} - q_{IA}{}'\right). \tag{6.34}$$

Two leptons, generated in two separate clusters, make up the initial state corresponding to the observer's time $t = -\infty$. By resetting the clock in s for each lepton, let this initial state correspond to $s = -\infty$ for each lepton. Similarly, let the observer's final state at $t = +\infty$ correspond to $s = +\infty$. The positions of the initial lepton's are measured in the spatial frame of the observer's choice, as discussed earlier. As a result, the spatial momentum and position coordinates are no longer co-linear and we shall assume

$$\langle \mathbf{q}_{IA} | \mathbf{p}_{IA} \rangle = (1/2\pi)^3 \exp(-i\mathbf{q}_{IA} \cdot \mathbf{p}_{IA}). \tag{6.35}$$

However, for the fourth components $p_{IA0}$ and $q_{IA0}$, the analogous expression still does not hold in general. Thus, the transforms (4.71) and (4.72) are replaced by

$$\begin{aligned} |p_{IA}, n_3\rangle = \frac{1}{(2\pi)^4} \int_{-\infty}^{\infty} d^3q_{IA} \left[ \Theta(p_{IA0}) \int_0^{\infty} dq_{IA0} + \Theta(-p_{IA0}) \int_{-\infty}^{0} dq_{IA0} \right] \\ \times e^{iq_{IA}p_{IA}} | q_{IA}, n_3\rangle ; \end{aligned} \tag{6.36}$$

$$\begin{aligned} |q_{IA}, n_3\rangle = \frac{1}{(2\pi)^4} \int_{-\infty}^{\infty} d^3p_{IA} \left[ \Theta(q_{IA0}) \int_0^{\infty} dp_{IA0} + \Theta(-q_{IA0}) \int_{-\infty}^{0} dp_{IA0} \right] \\ \times \delta(p_{IA}{}^2 - m_{n_3}{}^2)\, e^{-iq_{IA}p_{IA}} | p_{IA}, n_3\rangle ; \end{aligned} \tag{6.37}$$

In the strong-interaction model, quark confinement in the initial state $t = -\infty$ implied that the quarks were the confined in the final state as well. When we calculate the scattering amplitude for quark $Q_{IA}$ arising from interactions with the other three quarks, $Q_{JB}$, we shall see that the resulting scattering amplitude is the same for both constituents of a lepton, i.e., $Q_{I1}$ and $Q_{I3}$. Moreover, there is no scattering between constituents of a given lepton. Although this does not provide rigorous justification for the assumption that the lepton composed of $Q_{I1}$ and $Q_{I3}$ remains intact, we shall assume in fact that the lepton does so. In other words, with only intuitive justification, we shall assume the perturbative forces are not strong enough to disassociate the constituent quarks.

### D. Propagator formulation of spinless lepton-lepton scattering

We shall consider quark $Q_{IA}$ scattered by quark $Q_{JB}$. In the Interaction Picture, take $H_I^{IAJB}(s)$ to be

$$H_I^{IAJB}(s) = \frac{1}{4} e_{AB}{}^2 \langle p^f_{JB}, n^f_3, s' | \int_{-\infty}^{s'} ds'$$

$$\times \Big[ P_{IA}(s) P_{JB}(s') \Delta_{IJ}^{AB} (Q_{IA}(s), Q_{JB}(s'))$$

$$+ P_{IA}(s) \Delta_{IJ}^{AB} (Q_{IA}(s), Q_{JB}(s')) P_{JB}(s')\Big)$$

$$+\, P_{JB}(s')\Delta_{IJ}^{AB}\left(Q_{IA}(s),Q_{JB}(s')\right) P_{IA}(s)$$

$$+\,\Delta_{IJ}^{AB}\left(Q_{IA}(s),Q_{JB}(s')\right)\, P_{IA}(s)P_{JB}(s')\Big]\,\Big|p_{JB}^{i},n_{3}^{i},-S'\Big\rangle , \qquad (6.38)$$

where A=1, 3; B=1, 3; and the propagator $\Delta_{IJ}^{AB}\left(Q_{IA}(s),Q_{JB}(s')\right)$ remains to be specified. Self-interaction terms are excluded (I=J while A=B).

We can now express the first-order scattering amplitude for $Q_{IA}$ as

$$a_{IJ}^{ABfi}=\frac{1}{\mathcal{N}}\lim_{\varepsilon\to 0}\ \lim_{S\to\infty}\ \lim_{S'\to S}\ (-i)e_{AB}{}^{2}$$

$$\times\Big\langle\Phi_{IA}(S)\Big|\Big\langle\Phi_{JB}(S')\Big|\int_{-S}^{S}ds\int_{-S'}^{S'}ds'\exp(iH_{IAO}s)\exp(iH_{JBO}s')$$

$$\times\, S_{IJ}^{AB}\exp(-iH_{IAO}s)\exp(-iH_{JBO}s')\Big|\Phi_{JB}(-S')\Big\rangle\Big|\Phi_{IA}(-S)\Big\rangle , \qquad (6.39)$$

with the constant scattering operator $S_{IJ}^{AB}$ defined by

$$S_{IJ}^{AB}\equiv\frac{1}{4}\Big[P_{IA}P_{JB}\Delta_{IJ}^{AB}+P_{IA}\Delta_{IJ}^{AB}P_{JB}+P_{JB}\Delta_{IJ}^{AB}P_{IA}+\Delta_{IJ}^{AB}P_{IA}P_{JB}\Big], \qquad (6.40)$$

and the propagator is now the constant operator $\Delta_{IJ}^{AB}\left(Q_{IA},Q_{JB}\right)$ (still to be defined). The constant $\mathcal{N}$ is a normalizing constant.

Taking the limits in S and S′, we can write the scattering amplitude as

$$a_{IJ}^{ABfi}=\lim_{\varepsilon\to 0}(-i)\left(2\pi\right)^{2}e_{AB}{}^{2}\delta\left(H_{IAO}^{i}-H_{IAO}^{f}\right)\delta\left(H_{JBO}^{i}-H_{JBO}^{f}\right)$$

$$\times\,\delta\left(H_{IAO}^{i}\right)\delta\left(H_{JBO}^{i}\right)\mathcal{M}_{IJ}^{ABfi}, \qquad (6.41)$$

with $\mathcal{M}_{IJ}^{ABfi}$ defined as

$$\mathcal{M}_{IJ}^{ABfi}=\left(1/\mathcal{N}\right)\Big\langle p_{IA}^{f},p_{JB}^{f},n_{3}^{f}\Big|S_{IJ}^{AB}\Big|p_{IA}^{i},p_{JB}^{i},n_{3}^{i}\Big\rangle . \qquad (6.42)$$

Recall from Sec. IV that for ε finite and s tending towards infinity, the operators $a_3$ and $a_3^{\dagger}$ commute (see (4.39)). Also recall that the physical momenta of the final state are minus the eigenvalues of $P_{IA}$. It is then straightforward to calculate

$$\mathcal{M}_{IJ}^{ABfi}=\frac{1}{2\mathcal{N}}\ \delta\left(p_{IA}^{f}+p_{JB}^{f}-p_{IA}^{i}-p_{JB}^{i}\right)\ \left(p_{IA}^{i}+p_{IA}^{f}\right)\cdot\left(p_{JB}^{i}+p_{JB}^{f}\right)$$

$$\times\Big\langle\, p_{IAJB}^{f},n_{AB}^{f}\Big|\Delta_{IJ}^{AB}\Big|\, p_{IAJB}^{f},n_{AB}\Big\rangle , \qquad (6.43)$$

where

$$p_{IAJB} \equiv \frac{1}{2}\left(p_{IA} - p_{JB}\right); \quad n_{AB} \equiv \left|\left(-1\right)^{\frac{A-1}{2}} - \left(-1\right)^{\frac{B-1}{2}}\right| n_3 . \tag{6.44}$$

The next step in calculating the matrix element $M_{IJ}^{ABfi}$ is to insert the transform (6.36). However, we still have not specified the propagator $\Delta_{IJ}^{AB}$, i.e., we have not chosen a form for the vector potential **A**(**Q**). It is to that choice that we now turn our attention.

## E. A vector potential for the nonrelativistic collision of two point particles

Electrodynamics is defined through a vector potential $A_\mu$. We seek, however, a form of the vector potential which will include the weak interactions as well. In short, we seek a form for $A_\mu$ which will lead to the Feynman rules for spinless lepton-lepton electroweak scattering. Although it is tempting choose the vector potential of the Wheeler-Feynman model, defined in terms of the quark coordinates, it is not difficult to demonstrate that this does not lead to Feynman rules except for zero-mass (photon) “exchange.”

For an alternative approach, let us re-examine an elementary problem of classical mechanics, namely the nonrelativistic collision of two point particles in one-dimension. The problem is first reduced to the equivalent single particle in a potential which depends on position. It is customary to assume the potential is a step function of infinite height. In this way, regardless of velocity, the incoming particle hits an impenetrable wall.

This assumption of an infinite step can be avoided, however, if we assume the potential is velocity dependent. Consider particles 1 and 2 to have equal mass and let their coordinates be the functions of time $x_1(t)$ and $x_2(t)$, respectively. Further, suppose that the initial conditions are $x_1(-\infty) = -\infty$, $x_2(-\infty) = +\infty$. Designate the initial momenta as $p_1^i$, and $p_2^i$. We now define a potential

$$V = -\left(p_2^i - p_1^i\right)\left(\dot{x}_2(t) - \dot{x}_1(t)\right)\,\Theta\left(x_2(t) - x_1(t)\right), \tag{6.45}$$

where the step function $\Theta(x)$ is defined by

$$\Theta(x) = \begin{cases} 0, & x < 0 \\ 1, & x > 0 \end{cases} . \tag{6.46}$$

Introduce the center-of-mass coordinates

$$X = \frac{1}{2}\left(x_2 + x_1\right), \qquad x = x_2 - x_1. \tag{6.47}$$

Then we can write

$$V = \dot{x}(t)\; A\left(x(t)\right), \tag{6.48}$$

where

$$A(x) = 2p^i\; \Theta\left(x(t)\right), \tag{6.49}$$

and we have set $p^i = \left(p_1^i - p_2^i\right)/2$.

The Lagrangian can thus be expressed as

$$L = \frac{M}{2}\dot{X}^2 + \frac{\mu}{2}\dot{x}^2 - \dot{x}A(x), \tag{6.50}$$

where M=2m and $\mu = m/2$. The conjugate momenta are

$$P = M\dot{X}, \qquad p = \mu\dot{x} - A. \tag{6.51}$$

Forming the Hamiltonian $H = P\dot{X} + p\dot{x} - L$, we obtain

$$H = \frac{1}{2M}P^2 + \frac{1}{2\mu}\left(p + A\right)^2. \tag{6.52}$$

The equations of motion $\dot{p} = -\partial H/\partial x$ and $\dot{x} = \partial H/\partial p$ yield

$$\dot{p} = -\frac{1}{\mu}\left(p + A\right)\frac{\partial A}{\partial x}, \qquad \dot{x} = \frac{1}{\mu}\left(p + A\right). \tag{6.53}$$

We obtain

$$\dot{p} = -\dot{x}\frac{\partial A}{\partial x} = -2p^i\dot{x}\;\delta(x). \tag{6.54}$$

Integrating (6.54) over t, we find

$$p^f - p^i = -2p^i\int_{-\infty}^{\infty}\dot{x}\delta(x)dt = -2p^i\int_{-\infty}^{\infty}\delta(t - t_0)dt = -2p^i, \tag{6.55}$$

where $t_0$ is such that $x(t_0)=0$. Thus, we obtain the expected answer for two impenetrable particles in a head-on collision, $p_1^i = p_2^f$, $p_2^i = p_1^f$.

The vector potential chosen in (6.49) is not unique. The result (6.55) can also be obtained by choosing the potential

$$A(x) = 2p^i\;\Theta(-x^2). \tag{6.56}$$

We can form a relativistic generalization of (6.56) that yields the Wheeler-Feynman potential by defining a four dimensional potential

$$A_\mu [x(s)] = C\sqrt{(p^{i})^2}\,\left[\dot{x}_\mu(s_0)/[-2x(s_0)\cdot\dot{x}(s_0)]\right]\,\Theta[-x^2(s)]. \qquad (6.57)$$

C is a constant with dimension of length, and $s_0$ is such that $x^2(s_0)=0$. Repeating steps similar to the one-dimensional problem, we find

$$A_\mu [x(s)] = 2C\sqrt{(p^{i})^2}\int_{-\infty}^{s}\dot{x}_\mu(s')\,\delta[-x^2(s')]\,ds', \qquad (6.58)$$

which can be compared to the Wheeler-Feynman potential (6.3). However, as we stated earlier, this potential does not give rise to Feynman's rules when applied to the case of non-zero mass "exchange." So, instead of the analogue of the Wheeler-Feynman potential, we shall generalize the quantized version of the simpler expression (6.49).

The quantized Hamiltonian in the center-of-mass system is

$$H = \frac{1}{2\mu}\left[\mathbf{p}^2 + \mathbf{pA} + \mathbf{Ap} + \mathbf{A}^2\right], \qquad (6.59)$$

with

$$A = 2p^{i}\Theta(x). \qquad (6.60)$$

For constant H, the Schrödinger equation is given by

$$H\Psi[x(t)] = E\Psi[x(t)] = \frac{1}{2\mu}[p^{i}]^2\Psi[x(t)] \qquad (6.61)$$

This equation can be expressed as

$$\frac{\partial^2\Psi(x)}{\partial x^2} = -2ip^1\delta(x)\Psi(x) - 4ip^{i}\Theta(x)\frac{\partial\Psi(x)}{\partial x} + [p^{i}]^2\left[4\Theta(x)-1\right]\Psi(x). \qquad (6.62)$$

When x<0, this yields the free-particle wave equation

$$\frac{\partial^2\Psi(x)}{\partial x^2} = -[p^{i}]^2\Psi(x). \qquad (6.63)$$

When x>0, the equation becomes

$$\frac{\partial^2\Psi(x)}{\partial x^2} = -4ip^{i}\frac{\partial\Psi(x)}{\partial x} + 3[p^{i}]^2\Psi(x). \qquad (6.64)$$

We can assume the wave function $\Psi(x)$ is continuous at x=0, but not its derivative with respect to x. Let us integrate the general wave equation (6.62) over an infinitesimal range about x=0. This yields

$$-i\left.\frac{\partial\Psi}{\partial x}\right|_{x=0+}+i\left.\frac{\partial\Psi}{\partial x}\right|_{x=0-}=-2p^{i}\int_{-\infty}^{\infty}\delta(x)\Psi(x)dx=-2p^{i}\Psi(0). \qquad (6.65)$$

We retrieve the expected result $p^f=-p^i$.

Solving the wave equations (6.63) and (6.64) and applying the boundary condition at x=0 leads to the two independent solutions below:

$$\Psi(x)=\begin{cases} a\left[e^{ip^{i}x}-\frac{1}{2}e^{-ip^{i}x}\right], & x<0;\\ a\left[\frac{1}{2}e^{ip^{i}x}\right], & x>0;\\ b\left[e^{ip^{i}x}-\frac{2}{3}e^{-ip^{i}x}\right], & x<0;\\ b\left[\frac{1}{3}e^{i3p^{i}x}\right], & x>0. \end{cases} \qquad (6.66)$$

Return now to the equation for the discontinuity in $\Psi(x)$ given in (6.65). In terms of bra and ket notation, it becomes

$$\langle x=0_{+}|p|\Psi\rangle-\langle x=0_{-}|p|\Psi\rangle=-2p^{i}\int_{-\infty}^{\infty}\langle 0|x'\rangle\langle x'|\Psi\rangle dx'=-2p^{i}\langle 0|\Psi\rangle. \qquad (6.67)$$

In other words, the potential A(x) can be expressed as

$$A(x)=2p^{i}\int_{-\infty}^{x}\delta(x')dx'=2p^{i}\int_{-\infty}^{x}\langle 0|x'\rangle dx'=2p^{i}\int_{-\infty}^{t}\dot{x}(t')\langle 0|x(t')\rangle dt' \qquad (6.68)$$

It is this form that we shall generalize to four-dimensional space for the calculation of the lepton-lepton scattering amplitude.

## F. Generalizing the collision problem to four dimensions: Feynman propagator and Feynman rules

We are finally in a position to calculate the matrix element $\mathcal{M}^{ABfi}_{IJ}$. We shall evaluate it in the frame for which $p^{i}_{IAJB0}=p^{f}_{IAJB0}=0\ \forall\ \varepsilon$. Inserting the momentum transform (6.36) into the expression (6.43) for $\mathcal{M}^{ABfi}_{IJ}$, we find

$$M_{IJ}^{ABfi} = \frac{1}{2N}\,\delta\left(p_{IA}^{f} + p_{JB}^{f} - p_{IA}^{i} - p_{JB}^{i}\right)\left(p_{IA}^{i} + p_{IA}^{f}\right)\cdot\left(p_{JB}^{i} + p_{JB}^{f}\right)$$

$$\times \frac{1}{(2\pi)^{4}} \int_{-\infty}^{\infty} d^{3}\vec{q}\left[\Theta\left(p_{IAJB0}^{i}\right)\int_{0}^{\infty} dq_{0} + \Theta\left(-p_{IAJB0}^{i}\right)\int_{-\infty}^{0} dq_{0}\right] e^{i\vec{q}\cdot\left(\vec{p}_{IAJB}^{f} - \vec{p}_{IAJB}^{i}\right)}\,\Delta\left(q, n_{AB}\right), \tag{6.69}$$

with

$$n_{AB} = \left|(-1)^{\frac{A-1}{2}} - (-1)^{\frac{B-1}{2}}\right| n_{3}. \tag{6.70}$$

In analogy to the one-dimensional collision problem discussed earlier, assume the propagator takes the form

$$\Delta\left(q, n_{AB}\right) = \left\langle q, n_{AB} \,\middle|\, 0, n_{AB}\right\rangle. \tag{6.71}$$

In operator form, we can express the propagator as

$$\Delta\left(q, n_{AB}\right) = \delta(q)\,\delta_{n_{AB} n_{AB}}. \tag{6.72}$$

Utilizing the transform (6.37), we obtain

$$\Delta\left(q, n_{AB}\right) = \frac{1}{(2\pi)^{4}} \int_{-\infty}^{\infty} d^{3}k\left[\Theta\left(q_{0}\right)\int_{0}^{\infty} dk_{0} + \Theta\left(-q_{0}\right)\int_{-\infty}^{0} dk_{0}\right]\delta\left(k^{2} - m_{n_{AB}}{}^{2}\right) e^{-iqk}, \tag{6.73}$$

with

$$m_{AB}{}^{2} = \frac{M}{2N}\left[\left|(-1)^{\frac{A-1}{2}} - (-1)^{\frac{B-1}{2}}\right| n_{3} + 2\right]. \tag{6.74}$$

Replace the delta function with

$$\delta\left(k^{2} - m_{n_{AB}}{}^{2}\right) \to -\frac{1}{2\pi i}\left\{\frac{1}{\left[k_{0} - \left(\omega_{k}^{AB} - i\varepsilon\right)\right]\left[k_{0} + \left(\omega_{k}^{AB} - i\varepsilon\right)\right]}\right.$$

$$\left. - \frac{1}{\left[k_{0} - \left(\omega_{k}^{AB} + i\varepsilon\right)\right]\left[k_{0} + \left(\omega_{k}^{AB} + i\varepsilon\right)\right]}\right\}, \tag{6.75}$$

where $\omega_{k}^{AB} = \sqrt{\vec{k}^{2} + m_{n_{AB}}{}^{2}}$. Now we can evaluate the integrals over $k_0$ by going to contour integrals in the complex $k_0$ plane. Consider first the case $q_0>0$, and choose the contour $C = C_1+C_2+C_3$, where $C_1$ is along the real axis from 0 to $\infty$, $C_2$ is the quarter circle in quadrant positive Re $k_0$ and negative Im $k_0$, and $C_3$ is along the imaginary axis from $-\infty$ to 0. Recall that the quark operator $q_{IA}$ is proportional to $1/\varepsilon$. Thus the contribution from $C_3$ vanishes. The contour C encloses a pole at $k_0 = \omega_k - i\varepsilon$, coming from the first term in (6.75).

In similar fashion, for $q_0$ negative, choose the analogous contour in the opposite quadrant encloses a pole at $k_0 = -\omega_k^n + i\varepsilon$. This pole also comes from the first term of (6.75). We obtain, finally,

$$\Delta(q, n_{AB}) = \frac{-i}{(2\pi)^3}\left[\Theta(q_0)\int_{-\infty}^{\infty}\frac{d^3k}{2\omega_k^{AB}}e^{-iqk} + \Theta(-q_0)\int_{-\infty}^{\infty}\frac{d^3k}{2\omega_k^{AB}}e^{iqk}\right]$$

$$= \int_{-\infty}^{\infty}\frac{d^4k}{(2\pi)^4}e^{-ikq}\frac{1}{k^2 - m_{n_{AB}}^{\;2} + i\varepsilon}. \tag{6.76}$$

Thus, $\Delta(q, n_{AB})$ is the Feynman propagator $\Delta_F(q, \mu^2)$ for the Klein Gordon equation.

Return now to the matrix element $M_{IJ}^{ABfi}$ in (6.69), and choose the normalization constant $N$ to be

$$N \equiv \int_0^{\infty} e^{-ix}dx, \tag{6.77}$$

we obtain

$$M_{IJ}^{ABfi} = N'\,\delta\left(p_{IA}^f + p_{JB}^f - p_{IA}^i - p_{JB}^i\right)\frac{\left(p_{IA}^i + p_{IA}^f\right)\cdot\left(p_{JB}^i + p_{JB}^f\right)}{\left(p_{IA}^f - p_{IA}^i\right)^2 + m_{n_{AB}}^{\;2}}, \tag{6.78}$$

where $N'$ is a new proportionality constant. $M_{IJ}^{ABfi}$ is characterized by three types of interactions:

(1) For $I = J$, $A \neq B$ (interactions between constituents within a lepton),

$$M_{II}^{ABfi} = N'\,\delta\left(p_{IA}^f - p_{IA}^i\right). \tag{6.79}$$

(2) For $I \neq J$, $A = B$ (electrodynamic interaction between like constituents in different leptons; zero-mass "exchange"),

$$M_{IJ}^{AAfi} = N'\,\delta\left(p_{IA}^f + p_{JA}^f - p_{IA}^i - p_{JA}^i\right)\frac{\left(p_{IA}^i + p_{IA}^f\right)\cdot\left(p_{JA}^i + p_{JA}^f\right)}{\left(p_{IA}^f - p_{IA}^i\right)^2}. \tag{6.80}$$

(3) For $I \neq J$, $A \neq B$ (weak interaction between unlike constituents in different leptons; massive particle “exchange”),

$$M_{IJ}^{ABfi} = N\ N\,\delta\left(p^f_{IA} + p^f_{JB} - p^i_{IA} - p^i_{JB}\right) \frac{\left(p^i_{IA} + p^f_{IA}\right)\cdot\left(p^i_{JB} + p^f_{JB}\right)}{\left(p^f_{IA} - p^i_{IA}\right)^2 + m_W{}^2}, \qquad (6.81)$$

with

$$m_W{}^2 = \frac{M}{N}\left(n_3 + 1\right) = m_{n_3}{}^2 + \frac{M}{2N} n_3 . \qquad (6.81)$$

## VII. INTERNAL SPIN VARIABLES AND SUPERSYMMETRY

In Sec. III, a Lagrangian was introduced which is invariant under the interchange of lepton and hadron quarks, i.e., under the simultaneous interchange $Q_{I1} \leftrightarrow Q_{I2}$, $Q_{I3} \leftrightarrow Q_{I4}$ . This Lagrangian was later modified in Sec. IV by the inclusion of the parameter $i\varepsilon$. In terms of the coordinates $x_{IA}$, the new system is similarly invariant under the simultaneous transformations $x_{I1} \leftrightarrow x_{I2}$, $x_{I3} \leftrightarrow x_{I4}$. However, the $x_{IA}$ were not taken to be the quark coordinates, but were transformed to yield the defined set of quark coordinates $Q_{IA}$. The Lagrangian in terms of the $Q_{IA}$ is not invariant under the interchange $Q_{I1} \leftrightarrow Q_{I2}$, $Q_{I3} \leftrightarrow Q_{I4}$. Nevertheless, the lepton and hadron composites retain the same mass spectrum.

The definition of supersymmetry rests on the concept of internal spin, i.e., on systems composed of fermions and bosons. Although various forms of supersymmetry have been proposed, all are based on the supposition that the system displays the same mass spectrum for fermions and bosons. Although the inclusion of spin is beyond the scope of the model presented in this paper, we shall give a brief and nonrigorous argument of how its inclusion can lead to half-integer spin for the lepton composites and integer spin for the hadron composites (it should be kept in mind that the “leptons” and “hadrons” of this model consist of pairs of quarks). That is, the leptons are fermions and the hadrons are bosons, and the system displays a form of supersymmetry.

The harmonic-oscillator solutions in the asymptotic regions in s produce the composite particles, the leptons and hadrons. More specifically, let us consider the case $s \to +\infty$. Asymptotic quark momenta are obtained from (4.25), (4.29), and (4.37), and in the limit $s \to +\infty$, $\varepsilon \to 0$, and, in the Schrödinger Picture, can be taken to be

$$P_{IA} = \sqrt{M/2}\left[a_{IA} \pm \left(1/2\sqrt{N}\right) i \left(a_3^\dagger - a_3\right)\right], \quad A=1,\ 3 \ \text{(lepton quarks)}; \qquad (7.1)$$

$$P_{IA} = \sqrt{M/2}\left[\left(a_{IA} + \left(1/\sqrt{2N}\right)a_2\right) \pm i\left(1/2\sqrt{N}\right)\left(a_4^{\dagger} - a_4\right)\right],\quad A=2,\ 4\ \text{(hadron quarks)}. \tag{7.2}$$

As we shall see later, it is the "extra" term $a_2$ in the hadron quark vectors that leads to integral spin for the hadron composites. In order to introduce spin variables corresponding to the space-time variables

$$a_{IA},\ \left(1/\sqrt{2N}\right)a_2,\ a_A,\ \text{and}\ a_A^{\dagger}, \tag{7.3}$$

a second phase space is added to the phase space of these quark coordinates. We shall first consider the oscillator operators $a_A$ and $a_A^{\dagger}$. Following Casalbuoni,[33] we add the "fermion" creation and annihilation operators $a_A^F$ and $a_A^{F\,\dagger}$ to the Hamiltonian symmetrically to the product $a_A^{\dagger}a_A$. These fermion operators obey anticommutation relations

$$\left\{a_{A\mu}^F, a_{A\nu}^{F\dagger}\right\} = -g_{\mu\nu},\quad A=3,\ 4. \tag{7.4}$$

The eigenstates of the Hamiltonian can be taken to be simultaneous eigenstates of the operators $N_A$ and $\overline{N}_A$ defined below:

$$N_A \equiv -\left(a_A^{\dagger}a_A + a_A^{F\dagger}a_A^F\right),$$

$$\overline{N}_A \equiv a_A^{\dagger}a_A^F + a_A^{F\dagger}a_A,\quad A=3,\ 4. \tag{7.5}$$

Note that $\overline{N}_A^{\,2} = -N_A$.

For the spin variables corresponding to the coordinates $a_{IA}$ and $\left(1/\sqrt{2N}\right)a_2$, we follow Berezin and Marinov[34] and introduce spin phase space operators $\xi_{IA\mu}$, $\xi_{IA5}$, $\xi_{2\mu}$, $\xi_{25}$. These obey anticommutation relations

$$\left\{\xi_{IA\mu}, \xi_{IA\nu}\right\} = -g_{\mu\nu},\qquad \left\{\xi_{IA5}, \xi_{IA5}\right\} = 1;$$

$$\left\{\xi_{2\mu}, \xi_{2\nu}\right\} = -g_{\mu\nu},\qquad \left\{\xi_{25}, \xi_{25}\right\} = 1. \tag{7.6}$$

The two sets of generators each produce a Clifford algebra $C_5$, and are represented by Dirac gamma matrices as follows:

$$\xi_{IA\mu} = \frac{1}{\sqrt{2}}\gamma_{IA5}\gamma_{IA\mu},\quad \xi_{IA5} = \frac{1}{\sqrt{2}}\gamma_{IA5};$$

$$\xi_{2\mu} = \frac{1}{\sqrt{2}}\gamma_{25}\gamma_{2\mu},\quad \xi_{25} = \frac{1}{\sqrt{2}}\gamma_{25}; \tag{7.7}$$

with

$$\{\gamma_{IA\mu},\gamma_{IA\nu}\} = 2g_{\mu\nu}, \qquad \gamma_{IA5} = i\gamma_{IA0}\gamma_{IA1}\gamma_{IA2}\gamma_{IA3};$$

$$\{\gamma_{2\mu},\gamma_{2\nu}\} = 2g_{\mu\nu}, \qquad \gamma_{25} = i\gamma_{20}\gamma_{21}\gamma_{22}\gamma_{23}. \tag{7.8}$$

We now define a lepton quark spin variable $\Xi_{IA\mu}$ to correspond to $P_{IA\mu}$, choosing an expression analogous to that used by Ramond[35] to describe hadrons of half-integer spin:

$$\Xi_{IA\mu} = \xi_{IA\mu} + \left(1/2\sqrt{N}\right) i \left(a^{F\dagger}_{3\mu} - a^{F}_{3\mu}\right), \quad A = 1,\ 3. \tag{7.9}$$

It follows that

$$\{\Xi_{IA\mu},\Xi_{IA\nu}\} = -\,g_{\mu\nu}. \tag{7.10}$$

Suppose now that the Klein-Gordon constraint (4.10) is replaced by the constraint

$$\Xi_{IA}\cdot P_{IA} \approx 0, \quad \text{as } s \to +0. \tag{7.11}$$

Then, the Klein-Gordon constraint follows, or

$$P_{IA}{}^{2} \approx 0, \quad \text{as } s \to 0. \tag{7.12}$$

Since $\Xi_{IA}\cdot P_{IA}$ includes terms which change the eigenvalues of $N_3$ and $\overline{N}_3$, (7.11) yields a set of constraints in addition to those implied by (7.12). Using a spinor representation, we can express these as

$$\left(\gamma_{IA}\cdot a_{IA} - \overline{N}_3\right) u_{IA}(a_{IA}) \left|a_{IA}, N_3, \overline{N}_3\right\rangle = 0,$$

$$a^{F}_{3}\cdot a_3 \left|a_{IA}, N_3, \overline{N}_3\right\rangle = 0, \qquad \left(\gamma_{IA}\cdot a_3 - a_{IA}\cdot a_3\right) u_{IA}(a_{IA}) \left|a_{IA}, N_3, \overline{N}_3\right\rangle. \tag{7.13}$$

where $u_{IA}$ is a four-component spinor.

The first of these constraint equations represents a generalized Dirac equation and implies that the corresponding lepton composite has half-integer spin. The same equation yields the composite mass spectrum

$$a_{IA}{}^{2} = N_3. \tag{7.14}$$

There is, of course, a degeneracy in spin.

In similar fashion, we define a hadron quark spin operator for hadron quarks:

$$\Xi_{IA\mu} = \xi_{IA\mu} + \xi_{2\mu} + i\left(1/2\sqrt{N}\right)\left(a^{F\dagger}_{4} - a^{F}_{4}\right). \tag{7.15}$$

Following arguments analogous to the lepton case, we find that the hadron composites are characterized by integral spin, but have the same mass spectrum.

## VIII. SUMMARY AND DISCUSSION

Although not all of the model's predictions were explored here, its simplicity certainly limits experimental predictions to a qualitative nature. Nevertheless, the groundwork has been laid for future exploration. For example, just as spontaneous symmetry breaking in the model leads to different behaviors for leptons and hadrons, the addition of other broken internal symmetries may yield mass spectra and amplitudes in closer agreement with experiment. Mechanisms are also in place to examine inelastic scattering, through both more complex quark clusters and the introduction of perturbative Hamiltonians that don't commute with $n_A$.

The model is characterized by several changes in basic concepts, some of which have been examined in the past and rejected. In the present paper, it is the confluence of these changes that allows a self-consistent and physically realistic model. Taken alone and applied to earlier particle ontologies, any of these changes, i.e., a relativistic particle ontology, the evolution parameter s, quark constituents of hadrons and leptons, or the inclusion of $\varepsilon$ (H Ö $H_0 + H_I$ and no background frame), leads to difficulties.

Of particular interest is the infinitesimal parameter $\varepsilon$. Although its role is somewhat obscure and needs to be better understood, it is deeply connected to the model's quantum field-like and string-like behavior. In other words, the parameter $\varepsilon$ suggests that there exists a mathematical structure which allows fields and strings to be simulated by point particles. If one accepts this premise, it becomes clear that leptons as well as hadrons must consist of quarks (as described in this paper), in order to simulate fields associated with both types of particles.

Regardless of future progress of the model presented, it raises interesting new possibilities. In addition to the above, these include the true unification of forces based on lepton and hadron quark constituents, the extension of a particle ontology to include gravity, and the role of fundamental harmonic-oscillator quark "strings" in describing the other forces of electroweak interactions and gravity.